\documentclass[aps,twocolumn,pra,showpacs,amsmath,superscriptaddress,amssymb,longbibliography,showkeys,10pt]{revtex4-1}
\usepackage{graphicx}
\usepackage{epstopdf}
\usepackage[normalem]{ulem}
\usepackage{braket}
\usepackage{hyperref}
\usepackage[dvipsnames]{xcolor}
\usepackage{ulem}
\newcommand{\Expc}{\mathbb{E}}
\newcommand{\expc}[1]{\left\langle #1\right\rangle}
\allowdisplaybreaks

\begin{document}

\title{Telling different unravelings apart via nonlinear quantum-trajectory averages}

\newcommand{\ICFO}
{ICFO -- Institut de Ci\`encies Fot\`oniques, The Barcelona Institute of Science and Technology, 08860 Castelldefels (Barcelona), Spain}

\newcommand{\ICREA}{ICREA -- Instituci\'{o} Catalana de Recerca i Estudis Avan{\c{c}}ats, 08010 Barcelona, Spain}

\author{Eloy Pi\~nol}
 \thanks{These two authors contributed equally}
\affiliation{\ICFO}

\author{Th. K. Mavrogordatos $^*$}
\email{themis.mavrogordatos@fysik.su.se}
\affiliation{\ICFO}
 \affiliation{Department of Physics, AlbaNova University Center, SE 106 91, Stockholm, Sweden}

\author{Dustin~Keys}
\affiliation{Department of Mathematics, The University of Arizona Tucson, AZ 85721-0089 USA}

\author{Romain Veyron}
\affiliation{\ICFO}

\author{Piotr Sierant}
\affiliation{\ICFO}

\author{Miguel Angel Garc\'ia-March}
\affiliation{Instituto Universitario de  Matem\'atica Pura y Aplicada, Universitat Polit\`ecnica de Val\`encia, Camino de Vera, s/n, 46022 Valencia, Spain}

\author{Samuele~Grandi}
\affiliation{\ICFO}

\author{Morgan W. Mitchell}
\affiliation{\ICFO}
\affiliation{\ICREA}

\author{Jan Wehr}
\affiliation{Department of Mathematics, The University of Arizona Tucson, AZ 85721-0089 USA}

\author{Maciej Lewenstein}
\email{maciej.lewenstein@icfo.es}
\affiliation{\ICFO}
\affiliation{\ICREA}


\definecolor{mygreen}{rgb}{0,0.5,0}
\definecolor{mygrey}{rgb}{0.5,0.5,0.5}
\definecolor{myred}{rgb}{0.75,0,0}
\definecolor{myblue}{rgb}{0,0,0.75}
\definecolor{mymagenta}{cmyk}{0,1,0,0.12}
\definecolor{mycyan}{cmyk}{1,0,0,0.12}
\definecolor{myorange}{rgb}{1,0.5,0}
\definecolor{myviolet}{rgb}{0.5,0.0,0.75}
\definecolor{mybrown}{cmyk}{0,0.50,1,0.41}


\date{\today}

\begin{abstract}

The Gorini-Kossakowski-Sudarshan-Lindblad master equation (ME) governs the density matrix of open quantum systems (OQSs). When an OQS is subjected to weak continuous measurement, its state evolves as a stochastic quantum trajectory, whose statistical average solves the ME. The ensemble of such trajectories is termed an unraveling of the ME. We propose a method to operationally distinguish unravelings produced by the same ME in different measurement scenarios, using nonlinear averages of observables over trajectories. We apply the method to the paradigmatic quantum nonlinear system of resonance fluorescence in a two-level atom. We compare the Poisson-type unraveling, induced by direct detection of photons scattered from the two-level emitter, and the Wiener-type unraveling, induced by phase-sensitive detection of the emitted field. We show that a quantum-trajectory-averaged variance is able to distinguish these measurement scenarios. We evaluate the performance of the method, which can be readily extended to more complex OQSs, under a range of realistic experimental conditions.
\end{abstract}


\maketitle

{\it Introduction.}---Quantum systems interacting with Markovian environments are ubiquitous in the physical sciences. A main tool for studying their dynamics is the deterministic Gorini-Kossakowski-Sudarshan-Lindblad (GKSL) master equation (ME)~\cite{GKS76,Lind76}. This specifies the time evolution of the density matrix $\rho(t)$ as the system experiences both coherent and incoherent processes, with the latter involving leakage of state information to the environment~\cite{BP99,HarocheBook,BP02,OQS206}. Despite its generality and wide use, the GKSL ME does not fully describe the quantum dynamics when the environment includes measurement devices, which convert a portion of the leaked information to usable form. The temporal evolution conditioned on the measurement record $m$ defines a \textit{quantum trajectory}, in the ideal case expressed as the pure state $\rho_m(t) = \ket{\psi_m(t)}\bra{\psi_m(t)}$. The unraveling into pure states can be performed whenever the continuous evolution in the GKSL ME is governed by a commutator with a non-Hermitian Hamiltonian, and the initial state is pure. The resulting nonlinear Schr\"{o}dinger equation governing the evolution of normalized states in quantum-trajectory theory is a straightforward consequence of conditioning and is not considered to arise from any previously unknown inherent stochasticity~\cite{CarmichaelBook2}. Averaging $\rho_m(t)$ over the measurement record solves the GKSL ME~\cite{Accardi13}. The identification of the $\rho(t)$ with the ensemble average of $\rho_m(t)$ is an example of an \textit{unraveling} of the ME into a stochastic equation for the pure state $\rho_m(t)$~\cite{BKS1924, carmichael1993open, Gardiner92, DCM92, omnes1994, Blanchard1995, Korotkov99, Plenio1998, Hegerfeldt1996, Daley2014}.  Evidently, different measurement schemes correspond to different unravelings and lead to different ensembles of quantum trajectories. Unravelings thus provide information that is not available from the corresponding ME.

Quantities that are linear in the density matrix $\rho(t)$, such as averages of observables, are fully determined by the GKSL ME and, therefore, are independent of the choice of the unraveling dictated by a given measurement scheme. In this Letter, we develop nonlinear measures to differentiate unravelings, thus opening a way to access the physics beyond the ME. We demonstrate that evaluating an expectation value of a physical observable for a specified quantum trajectory $\rho_m(t)$, performing a nonlinear operation on the obtained result, and averaging the result over the measurement record, yields a quantity that allows for distinguishing different unravelings of the same GKSL ME. We focus on a paradigmatic open quantum system, the resonance fluorescence of a two-level atom, and consider unravelings corresponding to direct photodetection and to homodyne/heterodyne detection. We remark that while direct photodetection enjoys an obvious link with intensity correlations, measuring the spectrum of squeezing is inherently tied with homodyne detection, which provides a phase reference to a phase-dependent phenomenon~\cite{CarmichaelBook2}.

\begin{figure}
    \centering
    \includegraphics[width=0.48\textwidth]{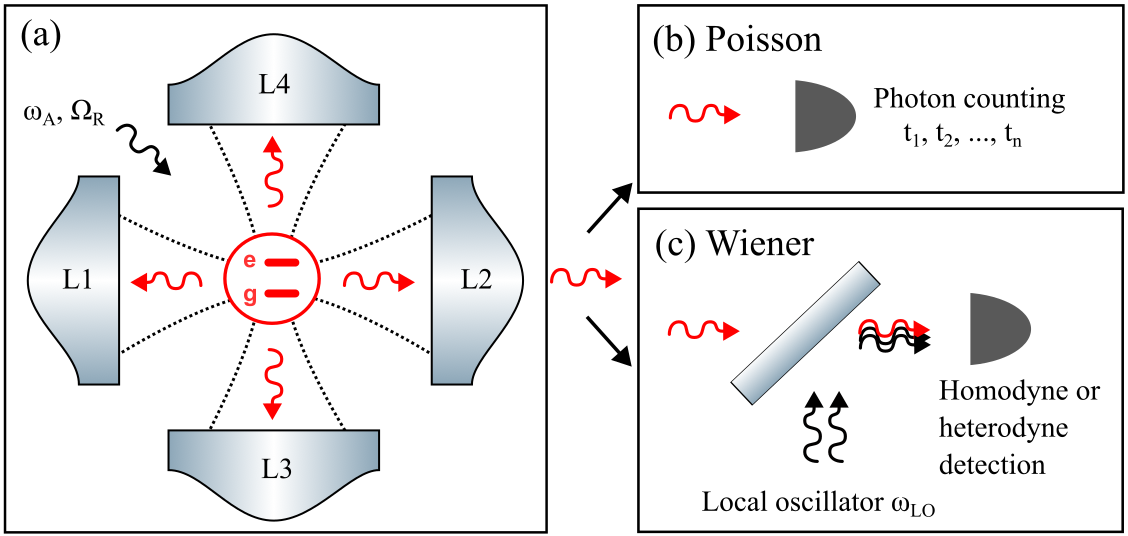}
    \caption{{\it Schematic representation of the two main unraveling schemes.} In this setup, the trapped two-state atom is illuminated in a Maltese-cross arrangement~\cite{Bruno2019,Markus2006} shown in {\bf (a)}. The output radiation escaping a particular lens is directed to either: {\bf (b)} a collection of avalanche photodetectors (APDs) producing a time-series of ``click'' events (Poisson-type unraveling); {\bf (c)} a mixer with a strong local oscillator field, to substantiate either a homodyne ($\omega_{\rm LO}=\omega_A$) or a heterodyne measurement ($|\omega_{\rm LO}-\omega_A| \gg \gamma$) scheme (Wiener-type unraveling).}
    \label{fig:scheme}
\end{figure}

{\it The unravelings.}---Electron shelving~\cite{Dehmelt,Dehmelt1} paved the way to the first observations of quantum jumps~\cite{Nagourney86, Sauter86, QuantumJumps}, followed by several atomic~\cite{Basche95, Peil99, Gleyzes07} and solid-state physics experiments~\cite{Neumann10, Vijay11, Minev2019}.
The theoretical description of these investigations dates back to early works~\cite{Cook, Kimble,Erber,Javanainen,Schenzle,Cohen,Cook88,Grochmalicki1989,Grochmalicki1989II, Hegerfeldt1992,Hegerfeldt1993,Hegerfeldt2009} that stimulated the development of quantum trajectory theory~\cite{KeysWehr,BarBel91, Brun2002,Bar06,Bel90,BelStas91,GPR90,GRW85, carmichael1993open}. Two unravelings of the GKSL ME that play a fundamental role in the understanding of the quantum trajectories are: (i) Poisson unraveling, related to direct photodetection and the so called quantum Monte Carlo wave function approach~\cite{DCM92,Davies76,SriDa,ZoWaMa, Dum92}; and (ii) Wiener-type unraveling (the quantum state diffusion model proposed by Gisin and Percival)~\cite{GP92, Carmichael1999}, relating conditional quantum dynamics to a continuous Wiener process~\cite{durrett19}. In the context of atomic physics experiments, the distinct unravelings correspond to different photodetection schemes~\cite{BP02}, see  Fig.~\ref{fig:scheme}. The Poisson unraveling is relevant for the direct photodetection experiments, while the continuous Wiener process arises in homodyne and heterodyne photodection schemes~\cite{carmichael1993open, Wiseman93}.

The disparity between the experimental setups is reflected in the different nature of quantum trajectories~\cite{Carmichael1999}. The Wiener process yields a continuous evolution of the system state $\ket{\psi(t)}$. In contrast, the acts of direct photodetection at times $t_1 < t_2 < \ldots < t_n$ collapse the conditional wavefunction. The resulting time evolution of $\ket{\psi(t)}$ is discontinuous and the final state at $t>t_n$ depends, in general, on a particular sequence of emission times $\ket{\psi(t)} =\ket{\psi_{t_1,\ldots,t_n}(t)}$. By collecting photon counting records, the experimenter effectively determines the quantum trajectory of the atom. The entanglement between the electromagnetic field and the atom is the key ingredient that allows for the inference of the atom's state based on the photodetection events~\cite{Carmichael1999, Nha2004}. In particular, Nha and Carmichael demonstrated that the degree of entanglement depends on how information in the environment is read~\cite{Nha2004}.

\begin{figure*}[!ht]
    \centering
    \includegraphics[width=\textwidth]{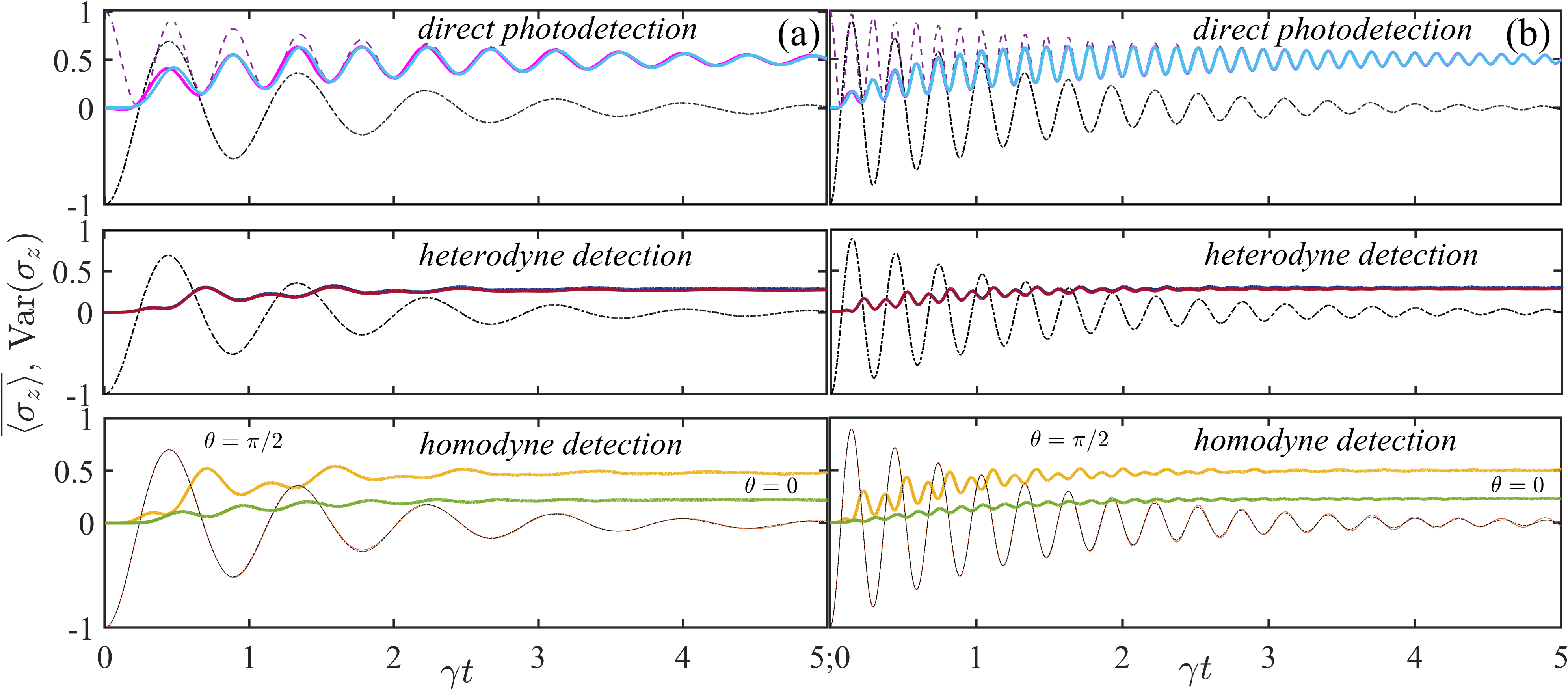}
    \caption{{\it Linear vs. nonlinear quantum-trajectory averages for three principal unravelings.} Monte-Carlo averages over $10^4$ realizations of the QTAV ${\rm Var}(\sigma_z)$ plotted against the dimensionless time $\gamma t$ for a Poisson-type unraveling (direct photodetection) and two Wiener-type unravelings (homodyne and heterodyne detection) as indicated in each panel, for {\bf (a)} $Y=10$ and {\bf (b)} $Y=30$, with the two-level atom initialized in its ground state. The oscillatory dot-dashed curves with alternating sign in all frames depict the average inversion $\overline{\langle \sigma_z(t) \rangle}$. In the uppermost panels of both frames, the pink and blue curves depict ${\rm Var(\sigma_z)}$ obtained from the perturbative treatment of the Dyson-series expansion to first order in $\gamma/\Omega$, and the moment-based equations, respectively. The latter results are indistinguishable from the Monte-Carlo simulations on the scale of the figure. The dashed curves (in purple) depict the asymptotic expression~\eqref{eq:asymptot}. For heterodyne detection, the QTAV obtained from the numerical simulations (in blue) is indistinguishable from the moment-based method results (in red). Homodyne detection is performed with the local-oscillator phase selected along the anti-squeezed and squeezed quadratures of the fluorescent field, at $\theta=0$ and $\pi/2$, respectively, corresponding to the same inversion average (brown curve overlapping with the dot-dashed).}
    \label{fig:diffunrav}
\end{figure*}

{\it Source Master Equation and linear averages.}---Our starting point is the GKSL ME of resonance fluorescence, governing the unconditional evolution of the reduced system density matrix $\rho$,
\begin{equation}\label{eq:ME}
\begin{aligned}
    \frac{d\rho}{dt}=\mathcal{L}\rho=&-i\tfrac{1}{2}\omega_A[\sigma_z, \rho] -i\Omega [e^{-i\omega_A t}\sigma_{+} +e^{i\omega_A t}\sigma_{-}, \rho]\\
    &+ \tfrac{\gamma}{2}(2\sigma_{-}\rho\sigma_{+} - \sigma_{+}\sigma_{-}\rho - \rho\sigma_{+}\sigma_{-}),
    \end{aligned}
\end{equation}
where we have neglected thermal excitation~\cite{SM}. In the ME~\eqref{eq:ME}, $\sigma_{+}, \sigma_{-}, \sigma_z$ are the raising, lowering and inversion operators (represented by Pauli matrices), respectively, for the two-level atom coherently driven by a resonant laser field of frequency $\omega_A$; $\Omega_R=2\Omega$ is the Rabi frequency at which the two-state atom periodically oscillates between its  ground and excited states, and $\gamma$ is the spontaneous emission rate. The solution of the corresponding optical Bloch equations yields the following expression for the average inversion when the atom is initialized in its ground state, 
\begin{equation}\label{eq:invME}
\overline{\langle \sigma_z(t)\rangle}\!=\!S_z \!\left[1 \!+\! Y^2 e^{-\left(3\gamma / 4 \right)t}\! \left(\!\cosh{\delta t}\! +\! \frac{\left(3\gamma / 4 \right)}{\delta} \sinh{\delta t}\!\right) \right],
\end{equation}
where $Y\equiv \sqrt{2}\Omega_R/\gamma$, $\delta\equiv \tfrac{\gamma}{4}\sqrt{1-8Y^2}$ and $S_z=-1/(1 + Y^2)$ is the steady-state inversion. Hereinafter, we denote by $\langle \cdot \rangle$ the quantum mechanical average over an individual realization. For strong driving ($Y \gg 1$) the average inversion exhibits damped oscillations at $\Omega_R$, relaxing to $0 + \mathcal{O}(\gamma^2/\Omega^2)$. Equation~\eqref{eq:invME} is an example of a typical linear average computed directly from the ME, against which our nonlinear averages are to be compared. We now describe the nonlinear averages. 

{\it Nonlinear averages beyond the density-matrix formalism.}---The idea underlying our approach is to perform a nonlinear operation on a quantum mechanical expectation value evaluated for an {\it individual} quantum trajectory prior to averaging of the result over the ensemble of quantum trajectories denoted by $\overline{(\circ)}$. A characteristic nonlinear average of our focus is the quantity ${\rm Var}(\sigma_z) \equiv \overline{\langle \sigma_z(t) \rangle^2} - [\overline{\langle \sigma_z(t) \rangle}]^2$, which we hereinafter call {\it quantum-trajectory-averaged variance} (QTAV). 

The results depicted in Fig.~\ref{fig:diffunrav} substantiate the pivotal influence of the environment when collecting records of a strongly driven two-state atom and taking a sum over a collection of them. The two principal unravelings are presented in their ability to produce an ostensibly  disparate ${\rm Var}(\sigma_z)$, while the corresponding average inversion remains unchanged. For the direct photodetection~\cite{Mollow1975, Cook1980, Cook1981}, corresponding to the Poisson-type unraveling of the ME, we obtain an exact expression for ${\rm Var}(\sigma_z)$ based on the waiting-time distribution~\cite{Carmichael1989, Grochmalicki1989}. For $\gamma t \gg 1$, the asymptotic expression for the variance, including first-order terms in $\gamma/\Omega$ of different frequencies, reads~\cite{SM}
\begin{equation}   \label{eq:asymptot}
\begin{aligned}
   {\rm Var}(\sigma_z) =& \tfrac{1}{2}\Big\{1 + e^{-\gamma t/2}\cos(4\Omega t) + \tfrac{\gamma}{8\Omega} e^{-\gamma t/2}[4\sin(4\Omega t)\\
   &-\sin(6\Omega t)-3\sin(2\Omega t)] + \mathcal{O}(\gamma^2/\Omega^2) \Big\}.
   \end{aligned}
\end{equation}

The first observation to be made from Eq.~\eqref{eq:asymptot} is that the amplitude of the dominant term (second term in the sum) to the QTAV -- revealing a frequency doubling with respect to the inversion -- is independent of $\Omega$. The variance ultimately relaxes to $1/2$, as we can see in both uppermost panels of frames (a) and (b). The asymptotic evolution to the steady state is in very good agreement with the exact Monte-Carlo simulations as well as with the perturbative treatment of the Dyson-series expansion for the variance~\cite{SM}, and the truncated hierarchy of moments produced from the adjoint Lindbladian. 

The time evolution of the QTAV is significantly altered, see the middle panels of Fig.~\ref{fig:diffunrav}, when one places a beam splitter and a local oscillator in the environment, and the fluorescent signal interferes with the latter before photodetection [Fig.~\ref{fig:scheme}(c)], corresponding to the heterodyne detection and exemplifying Wiener-type unraveling. The frequency doubling is also in evidence although the contrast in the oscillations is visibly suppressed. The light scattered by the two-level emitter is squeezed in the field quadrature that is in phase with the mean scattered field amplitude $\propto \langle\sigma_-(t)\rangle$~\cite{WallsZoller1981, Mandel1982}. The bottom panel in each frame shows that the QTAV responds differently to the detection of the squeezed {\it vs.} the direction of the antisqueezed quadrature of the fluorescent field, {\it i.e.}, along an axis perpendicular to the equator of the Bloch sphere where quantum fluctuations are redistributed among the quadratures.   
\begin{figure}
    \centering
    \includegraphics[width=0.5\textwidth]{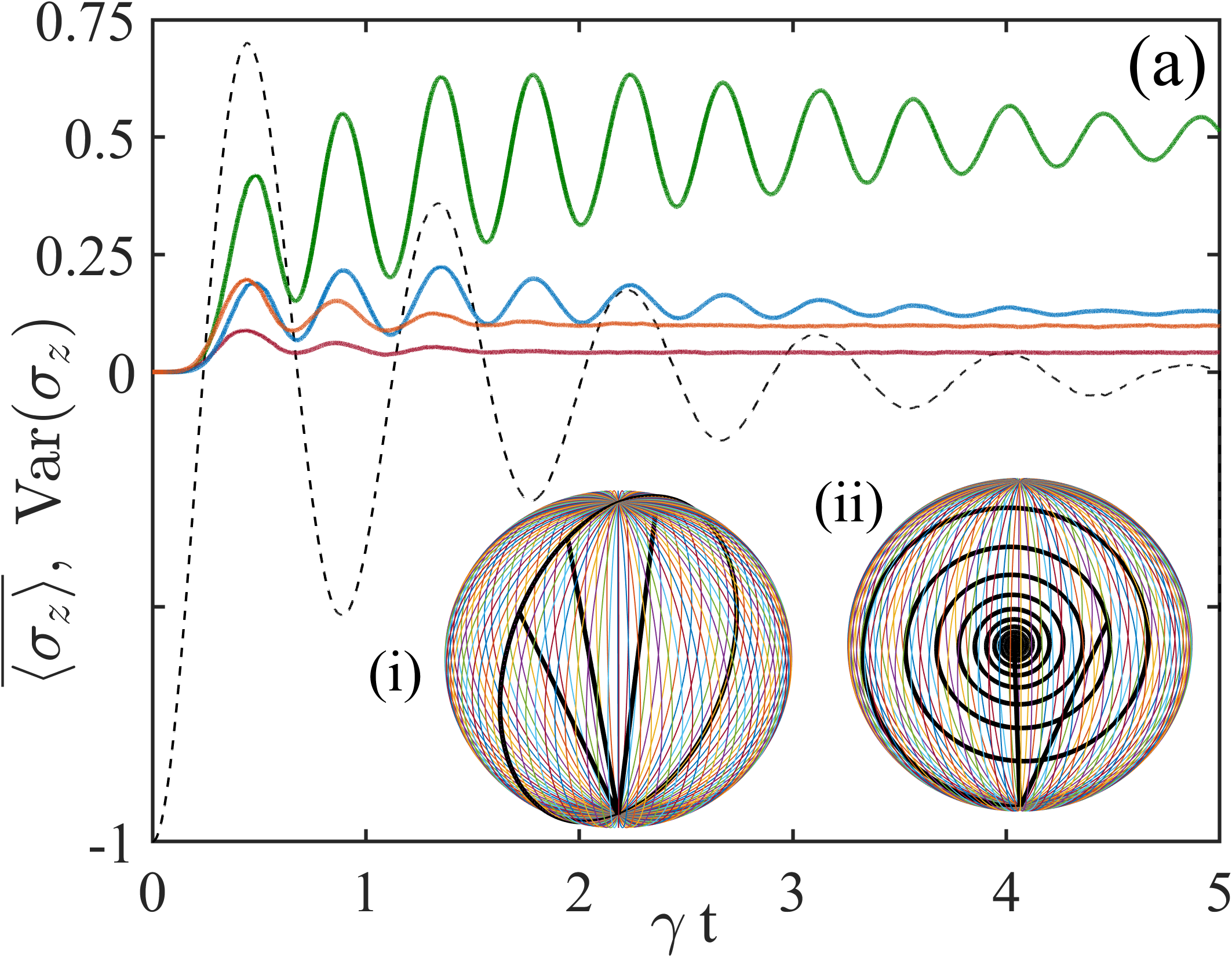}
    \includegraphics[width=0.5\textwidth]{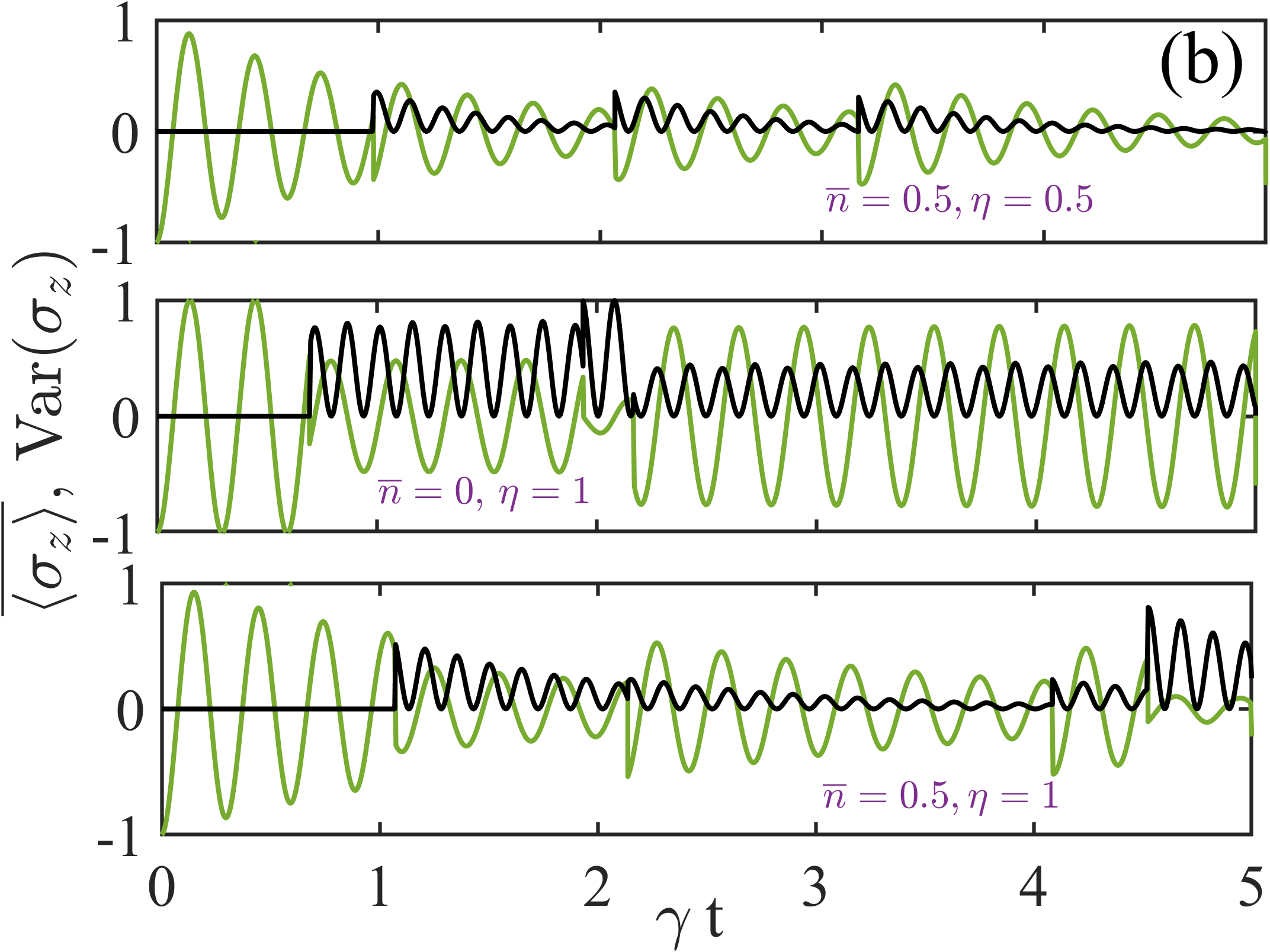}
    \caption{{\it Experimental limitations and decay of ``conditional'' coherence.} {\bf (a)} Monte-Carlo average over $10^4$ realizations of the QTAV ${\rm Var}(\sigma_z)$ obtained with $Y=10$, plotted against the dimensionless time $\gamma t$ for the ideal case  $\overline{n}=0, \eta=1$ (in green), $\overline{n}=0, \eta=0.5$ (in blue), $\overline{n}=1, \eta=1$ (in orange), and $\overline{n}=1, \eta=0.5$ (in red). The dashed curve depicts $\overline{\langle \sigma_z(t) \rangle}$ in the ideal case. The inset shows two sample trajectories in the Bloch sphere obtained with $Y=30$ and for $\overline{n}=0, \eta=1$ (i) and $\overline{n}=1, \eta=0.8$ (ii). {\bf (b)} Monte-Carlo average over {\it two} realizations for the atomic inversion, $\overline{\langle \sigma_z(t) \rangle}$ (in green) and  ${\rm Var}(\sigma_z)$ (in black) for the values of $\overline{n}, \eta$ indicated in each textbox.}
    \label{fig:explim}
\end{figure}

{\it Ensemble moments and adjoint Lindbladian.}---To provide some analytical grounding to the behavior
of the QTAV, we will delineate a method akin to the optical Bloch equations extended to account for 
the nonlinear averages. The contributions from the It\^o corrections to the ensemble moments can be found easily in the Heisenberg picture~\cite{KeysPhD}. Under the Poisson unraveling for an observable $A$~\cite{BarBel91} (we denote $\langle A(t) \rangle$ by $\langle A \rangle_t$):
	\begin{equation}
	d\expc{A}_t=\expc{\mathcal{L}^\dagger[A]}_t dt+\left(\dfrac{\expc{\sigma_+A\sigma_-}_{t-}}{\expc{\sigma_+\sigma_-}_{t-}}-\expc{A}_{t_{-}}\right)d\widetilde{N}(t),
	\label{HeisPoiss}
	\end{equation}
where $\widetilde{N}$ is the compensated Poisson process, $d\widetilde{N}=dN-\gamma\expc{\sigma_+\sigma_-}_tdt$, with a future pointing differential of expected value zero. This means that the ensemble average, here denoted by $\mathbb{E}$ for readability, is just
$$d\Expc\expc{A}_t=\Expc\expc{\mathcal{L}^\dagger[A]}_tdt,$$
which is the Heisenberg unraveling of the ME. It is useful to consider this equation for a Hilbert-Schmidt basis $X_i$ for the space of observables. Call the quantum expectation $x_i=\expc{X_i}$. Thus each observable $A$ has a corresponding vector $a$ such that $\expc{A}=(a,x)$. If we use the basis corresponding to the three Pauli matrices and the identity, all normalized in the Hilbert-Schmidt norm, then the vector for $A=\sigma_z$ is $a=(0,0,0,\sqrt{2})$. Since $a$ is just a constant vector, generally we have that $\Expc\expc{A}=\Expc(a,x)=(a,\Expc x)$ and the square of the quantum expectation becomes $\expc{A}^2=\sum_{ij}a_ia_jx_ix_j$ so that $\Expc\expc{A}^2=\sum_{ij}a_ia_j\Expc x_ix_j$ and so to see how the ensemble average of square of the quantum expectation evolves in time it is necessary to know how $\Expc x_ix_j$ evolves. In the Poisson case, we obtain
	\begin{align*}
		&d\Expc x_ix_j =\Expc\left(x_i\expc{\mathcal{L}^\dagger[X_j]}_tdt+\expc{\mathcal{L}^\dagger[X_i]}_tx_jdt\right) + \\
		&\Expc\left(\dfrac{\expc{\sigma_+X_i\sigma_-}_t}{\expc{\sigma_+\sigma_-}_t}-x_i\right)
		\left(\dfrac{\expc{\sigma_+X_j \sigma_-}_t}{\expc{\sigma_+\sigma_-}_t}-x_j\right)\\
        &\times\expc{\sigma_+\sigma_-}_tdt=\Expc\Big(x_i(u^j,x)+(u^i,x)x_j\\
        &+\dfrac{1}{(l,x)}\left((v^i,x_i)-x_i(l,x)\right)\left((v^j,x)-x_j(l,x)\right)\Big)dt,
	\end{align*}
using the fact that $d\widetilde{N}d\widetilde{N}=dN$ with the rate of the Poisson process being $\gamma\expc{\sigma_+\sigma_-}_t dt$ and in the last line using $u^j$ as the vector corresponding to $\mathcal{L}^\dagger[X_j]$, $l$ as $\sigma_+\sigma_-$, and $v^j$ as $\sigma_+X_j \sigma_-$. This is an ordinary differential equation which however does not close since it requires higher order moments such as $\Expc x_ix_jx_k$. Again, using the It\^o product rule we can calculate the equation for the higher order moments to obtain a system of ordinary differential equations which still do not close. We can repeat this procedure to arbitrarily high order but at some point we have to truncate. It can be shown that this truncation is linear in the moments, which allows us to solve the system using traditional methods of solving linear systems of ODEs. The Wiener case~\cite{BarBel91, GP92, gisin1997quantum, percival1998} can be similarly approximately solved by using the Heisenberg equation~\cite{KeysPhD}
	$$d\expc{A}_t\!=\!\expc{\mathcal{L}^\dagger[A]}_tdt\!+\!\left( \sqrt{\gamma}\expc{A(\sigma_--\expc{\sigma_-}_t)}\frac{dW(t)}{\sqrt{2}}\!+\!{\rm h.c.}\right),$$
where $W$ is a complex Wiener process. Solutions to these kind of truncated systems of equations for the two principal unravelings are depicted in Fig.~\ref{fig:diffunrav}, in very good agreement with the Dyson-expansion method and Monte-Carlo averages.

{\it Direct photodetection revisited and compromised.}---Having laid out an operational approach to distinguish the different unravelings, let us return to direct photodetection and discuss the most commonly encountered limitations in an actual experiment, where the density matrix {\it cannot} be unraveled into a pure-state ensemble, in which we would have a conditional wavefunction obeying a Schr\"odinger equation with a non-Hermitian Hamiltonian. This happens for a limited detector efficiency $\eta < 1$ and/or a surrounding bath with appreciable thermal excitation $\overline{n}$~\cite{SM}. 

Figure~\ref{fig:explim} testifies to the rapid degradation of ${\rm Var}(\sigma_z)$ as we move away from a pure-state description of the conditional dynamics. The QTAV responds to quantum jumps taking place in the course of individual realizations. This is evident from Fig.~\ref{fig:explim}(b) where ${\rm Var}(\sigma_z)$ remains zero until a spontaneous-emission event occurs in a pair of realizations. For an imperfect detector or for a thermally excited bath, the regression of fluctuations following a jump is damped. The decay concerns the coherent part of the evolution between spontaneous emissions, at a rate much faster than $\gamma$, for typical experimental parameters where $\eta \ll 1$. The inset of Fig.~\ref{fig:explim}(a) shows how sample trajectories spiral towards the center of the Bloch sphere, while individual jumps reset the evolution to the south pole. 

Since the intensity correlation of the scattered light reflects a nonexclusive probability of photocounting coincidences, the limited detector efficiency can be counterbalanced by increasing the number of photon ``clicks'' $N$ in the course of a long experimental run. Indeed, for the setup pictured in Fig.~\ref{fig:scheme}(a), we have concluded that the signal-to-noise ratio for $g^{(2)}(\gamma\tau \gg 1)$ in a window about one inverse of the coherence time scales with $\sqrt{N}$~\cite{SM}. This allows the determination of ${\rm Var}(\sigma_z)$ from single realizations as low as $10^{-3}$, which is the order of magnitude Monte-Carlo simulations indicate for $\eta\lesssim 0.05$. This order of magnitude can be increased using high numerical aperture collection systems~\cite{Bruno2019, Natarajan_2012} and efficient single-photon detectors~\cite{Chou2017}.
 
{\it Conclusions and outlook.}---In summary, we have expanded upon the fundamental concept of the variance in quantum mechanics going beyond the conventional density-matrix formulation. The different environments devised to collect the output of an open quantum system show up in a markedly different response of a quantity where nonlinear operations are performed to {\it individual} realizations prior to averaging over their ensemble. This is in contrast to linear observable averages where the complementary measurement strategies all abide by the predictions of the GKSL equation, and multi-time correlations -- such as the intensity correlation function -- are obtained via the quantum regression formula. Following our strategy, we need to set the initial point for two copies of the system (here a ground-state reset for direct photodetection) and then post-select the trajectories in such a way that the photocounting record is the same with satisfactory accuracy. {\it Ergo}, one gains, in principle, the ability to characterize the experiment’s power to collapse the wavefunction and add information to the memory carried by a state conditioned on all events that have taken place along a single trajectory. 

This ability allows for an experimentally oriented test of the objective quantum state assumption via an EPR steering inequality~\cite{Jones2007, Wiseman2012}; direct photodetection has been argued to be `more quantum' than its Wiener-type counterparts~\cite{Jones2007, Wiseman2012, Daryanoosh_2014}. While we have focused on two primary unravelings, the possibilities are in fact endless as different output channels open up. For example, Barchielli and Gregoratti have used a measurement-based feedback protocol to assess the non-Markovian evolution of a coherently driven and continuously monitored two-level atom. The included delay has experimental consequences, modifying the Mandel Q parameter alongside the spectrum of the emitted light~\cite{BarchielliGregoratti2012}. In general, non-Markov open systems cannot be given a trajectory interpretation built around measured outputs, as measuring non-Markov environments interferes with the reduced system dynamics. One sub-class of trivially non-Markov open systems has been recently approached using trajectories~\cite{Arranz2021}. It would be interesting to extend our considerations to non-Markovian evolution case~\cite{BarchielliGregoratti2012}. Finally, our conclusions are reflected by recent investigations of quantum many-body systems studied in the context of quantum computing and quantum simulation~\cite{Preskill2018quantumcomputingin, Altman21, fraxanet2022coming}. Quantum trajectories arising due to multiple measurements of the system's state, when analyzed by relevant (nonlinear) statistical measures such as entanglement entropy, exhibit phase transitions~\cite{skinner2019measurementinducedphase,li2018quantumzenoeffect,li2019measurementdrivenentanglement,chan2019unitaryprojective,sierant2022universalbehaviorbeyond, Turkeshi20_2d, lunt2021measurementinducedcriticality, Sierant22_2d}, that are not evident in the average state~\cite{cao2019entanglementina, Piccitto2022, Kolodrubetz2021, Turkeshi21clicks} unless specifically tuned feedback mechanisms are employed~\cite{Sierant23control, buchhold2022revealingmeasurementinduced,iadecola2022dynamicalentanglementtransition,Piroli23triviality,odea2022entanglementandabsorbing,ravindranath2022entanglementsteeringin,ravindranath2023free,sierant2023entanglement}.

\appendix
\onecolumngrid 
\setcounter{equation}{0}
\setcounter{figure}{0} 
\renewcommand\thefigure{S\thesection.\arabic{figure}} 
\renewcommand{\theequation}{S\thesection.\arabic{equation}}
\section*{Supplementary Information}

In the supplementary material, we first detail the calculation of the quantum-trajectory-averaged variance ${\rm Var}(\sigma_z)$ for the Poisson-type unraveling, via the Dyson-series expansion of the conditional reduced system density operator. We derive exact and approximate results, the latter in the limits of strong and weak driving where asymptotic expressions can be obtained. Secondly, we expand on the generality of the moment-based method applicable to the two principal types of unraveling (Poisson and Wiener). We also connect the quadrature amplitude squeezing encountered in resonance fluorescence to ${\rm Var}(\sigma_z)$. Finally, we take into account experimental imperfections and discuss a strategy to determine nonlinear averages in an exemplary and pioneering system of a single trapped fluorescing $^{87}$Rb atom whose output radiation is collected by four partially transmitting mirrors in a Maltese-cross arrangement.

\subsection{Introduction: record making and complementarity}

The theory of quantum trajectories ultimately attempts to describe the energy exchange between light and atoms, given both the quantum and wave aspects of light. The exchange must be described in a background where the quantum indicates discontinuity while the wave indicates continuity, and quantum trajectories fit both aspects in an evolution over time~\cite{BKS1924}. They employ the random stochastic processes and a formal generalization of the quantum jump to account for coherence~\cite{Hegerfeldt2009, carmichael1993open}. Both event-enhanced quantum theory~\cite{Blanchard1995} and consistent histories~\cite{omnes1994} emphasize the need to attach meaningful time series of real numbers to a quantum evolution. The series are the records obtained in the scattering scenario of a quantum optical experiment.    

By producing photon counting records, we effectively define the environment by a particular idealization of what {\it might} lie in the path of the scattered field -- a perfectly absorbing boundary. Every photon scattered by the two-state atom is then used up making a record appropriate to this environment. Other idealized environments will produce different records, and disentangle the system and environment in different ways~\cite{Nha2004}. There are many different environments that might, in fact, be encountered by the scattered field, all consistent with the master equation (ME). Different environments correspond to mutually exclusive methods of record making, since every photon produces one and only one happening. Each idealized environment defines a self-consistent pure-state unraveling. This is how quantum-trajectory theory encounters Bohr's complementarity. Apart from direct photodetection, there is another particularly important way of making records. It introduces a beam splitter and a local oscillator into the environment, and after the beam splitter every photon is counted. The scheme of homodyne and heterodyne detection uses {\it interference} to unveil aspects of the scattering process associated with a wave amplitude and a spectrum~\cite{Carmichael1999}. In the sections that follow, we will visit these complementary unravelings of the ME of resonance fluorescence, when producing the single realizations which make the quantum-trajectory-averaged variance (QTAV).

\subsection{Resonance fluorescence and waiting-time distribution}
We work with the paradigmatic system of resonance fluorescence, comprising a coherently driven two level atom (whose ground and excited states are denoted by $\ket{\downarrow}$ and $\ket{\uparrow}$, respectively) immersed in the vacuum reservoir. The derivation of the  photoelectron counting distribution by Mollow~\cite{Mollow1975} and Cook~\cite{Cook1980, Cook1981} is based on a hierarchy of equations that yield the probabilities for finding $n$ photons in the multimode fluorescent field. These equations were then used in the analysis of quantum jumps~\cite{ZoWaMa}. 

In such a system, the trajectories themselves are Markovian  (as well as the averaged dynamics conforming to the Gorini-Kossakowski-Sudarshan-Lindblad (GKSL) equation), since the dynamical evolution is reset to the same state following a quantum jump. As usual, we denote by $\sigma_{-}$ and $\sigma_{+}$ the lowering and raising system operators, respectively, $\Omega$ is the Rabi frequency which is taken real without loss of generality, and $\gamma$ is the spontaneous decay rate. The conditional evolution of the system state under direct photodetection (with unit detector efficiency) is described by the ME~\cite{GKS76,Lind76}
\begin{equation}\label{eq:MEcond}
\dot\rho_c = \mathcal{L} \rho_c = (\ell + J) \rho_c ,
\end{equation}
where in the interaction picture we may write
\begin{equation}
\ell \rho_c = [1/(i\hbar)] [H, \rho_c] - \frac{\gamma}{2}\left(\sigma_{+} \sigma_{-} \rho_c + \rho_c     \sigma_{+} \sigma_{-} \right ) = [1/(i\hbar)] (H_{\rm eff} \rho_c - \rho_c H_{\rm eff}^\dag),
\end{equation}
in which $H_{\rm eff}$ is a non-Hermitian Hamiltonian.  It describes a continuous evolution of the system state with decreasing norm, between randomly occurring spontaneous emission events
\begin{equation}\label{eq:Heff}
H_{\rm eff} = \hbar\Omega (\sigma_{+} + \sigma_{-}) -i\hbar(\gamma/2)\sigma_{+}\sigma_{-}.
\end{equation}
The continuous evolution is interrupted by quantum jumps accounted for by the action of the super-operator
\begin{equation}
J\rho_c = \gamma \sigma_{-} \rho_c \sigma_{+}.
\end{equation}
The above jump superoperator projects the system state to the ground state ($\ket{\downarrow}$), captures the aftermath of a photon emission. The time $\tau$ lapsed (often called waited time) between successive emissions is governed by the waiting-time
distribution $w(\tau)$. This exclusive probability density function of the time intervals $\tau$ between two consecutive jumps, is given by the expression
\begin{equation}
w\left(\tau\right) = {\rm Tr}\left (J e^{\ell \tau} (\left|\downarrow\right\rangle \left\langle \downarrow \right|)\right) = \gamma |\langle \uparrow| e^{[1/(i\hbar)]H_{\rm eff} \tau} |\downarrow\rangle|^2 = \exp \left(-\frac{\gamma \tau}{2}\right)\frac{\gamma \Omega^2}{\mu^2}\sin^2{\left(\mu\tau \right)},
\end{equation}
with $\mu = \frac{1}{2}\sqrt{4\Omega^2-\left(\frac{\gamma}{2}\right)^2}$. The waiting-time distribution forms the basis of the quantum-trajectory description of resonance fluorescence in direct photodetection~\cite{Carmichael1989}, as we will see in the following sections. 
\subsection{Dyson expansion and quantum-trajectory formulation in direct photodetection}
The solution of the ME~\eqref{eq:MEcond} can be expressed by means of the Dyson expansion \cite{carmichael1993open,SriDa,ZoWaMa, Dum92} as follows
\begin{equation}\label{eq:Dyson_expansion}
\rho_c\left( t \right) = e^{\ell t} \rho_c\left( 0 \right) + \int_0^t e^{\ell \left( t - t_1 \right) } J e^{\ell t_1} \rho_c\left( 0 \right)\mathrm{d}t_1 + \int_0^t \int_0^{t_2} e^{\ell \left( t - t_2 \right) } J e^{\ell \left( t_2 - t_1 \right) } J e^{\ell t_1} \rho_c\left( 0 \right)\mathrm{d}t_2\mathrm{d}t_1 + ...
\end{equation}
This form of the solution is very useful to see what an average over all the possible different quantum trajectories is made of, i.e. track the conditioned evolution paths. For example, the first term of the RHS describes a trajectory with no jumps at all. The second term describes all the possible trajectories with one jump at any time instant during the evolution, and so on. In fact, assuming $\rho\left( 0 \right) = \left|\downarrow\right\rangle \left\langle \downarrow \right |$, this solution can be recast in the explicit form of an average
\begin{equation}
\rho_c\left( t \right) = \sum_{n=0}^\infty \int_0^t \int_0^{t_n}...\int_0^{t_2} dt_n ... dt_1 \frac{e^{\ell\left(t - t_n \right)} \left(\left|\downarrow\right\rangle \left\langle \downarrow \right |\right)}{p_0(t - t_n )} p_n (t, t_1, t_2, ..., t_n),
\end{equation}
where 
\begin{equation}
p_n (t, t_1, t_2, ..., t_n) =  p_0 (t - t_n) w(t_n - t_{n-1}) ... \; w(t_2 - t_1) w(t_1)
\end{equation}
is the exclusive probability density for realizing one particular trajectory with $n$ jumps at times $t_1, t_2, ..., t_n$ and no jumps between $t_n$ and $t$~\cite{BP02}. Here, 
\begin{equation}\label{eq:p0}
p_0 (t - t_n) ={\rm Tr}(e^{\ell\left(t - t_n \right)}\left|\downarrow\right\rangle \left\langle \downarrow \right |) = e^{- \frac{\gamma (t - t_n)}{2}}\left[\frac{\Omega^2}{\mu^2} - \frac{\gamma^2}{16\mu^2}\cos{\left(2\mu \left(t - t_n \right) \right)} + \frac{\gamma}{4\mu}\sin\left(2\mu \left(t - t_n \right) \right) \right]
\end{equation}
is the null measurement probability density, from the time $t_n$ when the last jump was recorded to the final time $t$.
\subsubsection{Ensemble average of nonlinear functions of quantum mechanical expected values}
Using the previously derived expressions, one can obtain the following formula for the ensemble average of the quantum mechanical expected value of an operator $O$:
\begin{equation}\label{eq:averageO}
\overline{\braket{O (t)}} = {\rm Tr}\left(O \rho_c\left (t \right) \right) = \sum_{n=0}^\infty \int_0^t \int_0^{t_n}...\int_0^{t_2} dt_n ... dt_1 \frac{{\rm Tr} \left(O e^{\ell\left(t - t_n \right)}\left|\downarrow\right\rangle \left\langle \downarrow \right |\right)}{p_0(t - t_n )} p_n (t, t_1, t_2, ..., t_n).
\end{equation}
Here, the overbar denotes the ensemble average over all the possible trajectories and the brackets for the quantum mechanical expected value of each one of them. It is the common average obtained from the ME. Based on Eq.~\eqref{eq:averageO} we can construct a nonlinear average where the single-trajectory quantum mechanical average is raised to some power, i.e., after effecting a nonlinear operation in a post-selection process
\begin{equation}\label{eq:wL}
\overline{\braket{O (t)}^m} = \sum_{n=0}^\infty \int_0^t \int_0^{t_n}...\int_0^{t_2} dt_n ... dt_1 O_{m} \left(t - t_n \right) w(t_n - t_{n-1}) ... \; w(t_2 - t_1) w(t_1),
\end{equation}
with
\begin{equation}\label{eq:Om}
O_{m}\left(t \right) = \frac{\left[{\rm Tr} \left(O e^{\ell t} \left( \left|\downarrow\right\rangle \left\langle \downarrow \right |\right) \right)\right]^m}{p_0^{m-1}(t)},
\end{equation}
a considerably simplified form given that after the last ($n$) jump the wavefunction has collapsed to $\ket{\downarrow}$ (the individual trajectory is Markovian).

We remark that Eq.~\eqref{eq:wL} is an expression which cannot be obtained from the ME without the quantum trajectories point of view of the system evolution. Note that the expression for $\overline{\braket{O (t)}^m}$ is just a sum over successive convolutions:
\begin{equation}
\overline{\braket{O (t)}^m} = \sum_{n=0}^\infty O_{m} (t - t_n) \ast \Bigl\{ w\left(t_n - t_{n-1} \right) ... \ast \bigl[ w\left(t_2 - t_1 \right) \ast w\left(t_1 \right) \bigr] \Bigr\}.
\end{equation}
Now, applying the Laplace transform (which we denote as an upper tilde), we obtain the following expression for the ensemble average of the nonlinear quantum mechanical expected value:
\begin{equation}
\widetilde{\overline{\braket{O (t)}^m}} = \sum_{n=0}^\infty \widetilde{O_{m}}\left(z \right) \widetilde{w}^n\left(z \right) = \frac{\widetilde{O_{m}}\left(z \right)}{1 - \widetilde{w}\left(z \right)}.
\end{equation}
\subsubsection{Characteristic examples of nonlinear averages obtained via the Dyson expansion}
To illustrate the method, we chose the Pauli operator $\sigma_z$ and as a nonlinear function of the quantum mechanical average we select the square, i.e., $m = 2$. In this case,
\begin{equation}
\sigma_{z_{\;m=2}} (t) = \frac{\left[{\rm Tr} \left(\sigma_z e^{\ell t} \left( \left|\downarrow\right\rangle \left\langle \downarrow \right |\right) \right)\right]^2}{p_0(t)},
\end{equation}
with the denominator given by Eq.~\eqref{eq:p0}. Working on the numerator, we obtain the exact expression:
\begin{equation}
  \Bigl({\rm Tr}\left(\sigma_z e^
{\ell t} \left(\left|\downarrow\right\rangle \left\langle \downarrow \right | \right) \right) \Bigr)^2=e^{-\gamma t} \left\{\frac{1}{2} + \frac{\gamma^2}{32\mu^2} + \frac{1}{2} \left[1 - \frac{\gamma^2}{16\mu^2} \right] \cos(4\mu t) + \frac{\gamma}{4\mu} \sin(4\mu t)\right\},  
\end{equation}
identifying a dominant oscillatory term of frequency $4\mu$. A frequency mixing, however, is bound to arise due to the denominator $p_0(t)$, albeit scaled by powers of $\gamma/\Omega$. 

The above observation brings us to the strong-driving limit, $Y \gg 1$, with $Y \equiv 2\sqrt{2}\Omega/\gamma$. Neglecting second-order terms in $(\gamma/\Omega)^2 \ll 1$ we write $\mu \approx \Omega$; taking the Laplace transform of $\sigma_{z_{\;m=2}} (t)$, multiplying by $1/[1-\tilde{w}(z)]$ and going back to the time domain yields the following approximate form:
\begin{equation}\label{eq:szm2}
\begin{split}
\overline{\braket{\sigma_z (t)}^2} &\approx \frac{1}{2} + \frac{1}{4}e^{-\frac{3}{4} \gamma t} \Bigl[ C_1 \cos{(C_\Omega t)} + C_2 \sin{(C_\Omega t)} \Bigr] + \\
&+ \frac{1}{4}e^{-\frac{\gamma}{2} t} \Bigl[ C_3 \cos{(4\Omega t)} + C_4 \sin{(4\Omega t)} + C_5 \cos{(6\Omega t)} + C_6 \sin{(6\Omega t)}\Bigr].
\end{split}
\end{equation}
The coefficients $C_1, ..., C_6$ depend nonlinearly on the ratio $\gamma/\Omega$, while $C_{\Omega} \approx 2\Omega$, in agreement with the occurrence of the pair of eigenvalues with real part $-3\gamma/4$ (distinct from the vertical line at $-\gamma/2$) shown in Fig.~\ref{fig:EigFig}. 

In the case of $\sigma_y$, we find that the form of the solution is the same, and only the coefficients $C_1-C_6$ change. On the other hand, for the case of $\sigma_x$, the solution has an easier form, but it is different from zero only in the presence of a finite detuning $\Delta$ between the laser drive and the atomic resonance.

\subsubsection{Asymptotic results for strong and weak drive in the long-time limit}

Let us now consider the long-time limit of the QTAV from the perspective of the final-value theorem in the Laplace Transform. We are still working under $Y \gg 1$ (with $\mu \approx \Omega$). We then obtain
\begin{equation}
   \Bigl({\rm Tr}\left(\sigma_z e^
{\ell t} \left(\left|\downarrow\right\rangle \left\langle \downarrow \right | \right) \right) \Bigr)^2 = \tfrac{1}{2} e^{-\gamma t}[1+\cos (4\Omega t)] + \mathcal{O}(\gamma/\Omega),
\end{equation}
while $p_0(t) \approx e^{-\gamma t/2} + \mathcal{O}(\gamma/\Omega)$. As we have seen in the previous section, this leaves us to leading order with a damped oscillatory term $ e^{-\gamma t/2} \cos^2 (2\Omega t) + \mathcal{O}(\gamma/\Omega)$, ``screened" by a convolution kernel, the inverse Laplace Transform of $[1-\tilde{w}(z)]^{-1}$. Applying the final-value theorem, where $[1-\tilde{w}(z)]^{-1} \approx (z+\gamma/2)/z$ for $z \sim \gamma$, yields $1/2 + \mathcal{O}(\gamma^2/\Omega^2)$ for the long-time limit of ${\rm Var}(\sigma_z)$. Given that $S_z \equiv \langle\sigma_z \rangle_{\rm ss}=0 + \mathcal{O}(\gamma^2/\Omega^2)$, the asymptotic expression for the time-evolving variance--including first-order terms in $\gamma/\Omega$ of different frequencies--becomes (for $t \gg \gamma^{-1}$)
\begin{equation}
    \boxed{{\rm Var}(\sigma_z)= \tfrac{1}{2}\left\{1 + e^{-\gamma t/2}\cos(4\Omega t) + \tfrac{\gamma}{8\Omega} e^{-\gamma t/2}[4\sin(4\Omega t)-\sin(6\Omega t)-\sin(2\Omega t)] - \tfrac{\gamma}{4\Omega} e^{-\gamma t/2}\sin(2\Omega t) \right\} + \mathcal{O}(\gamma^2/\Omega^2),}
\end{equation}
which makes Eq. (3) of the main text, one of our central results. Although limited in its applicability, is shows that all oscillatory terms decay with the same rate, $\gamma/2$, consistent with the eigenvalue distribution plotted in Fig.~\ref{fig:EigFig}. Comparing with Eq.~\eqref{eq:szm2}, we expect that the coefficient $C_3$ does not scale with $\gamma/\Omega$ for $Y \gg 1$ but approaches a constant value. Indeed, we find that $C_3 \approx 2$, while $C_4\sim 1/Y$. At the same time, the initial-value theorem gives ${\rm Var}(\sigma_z)(t\to 0^+)=0 + \mathcal{O}(\gamma^2/\Omega^2)$, which we have also numerically confirmed (see Fig. 2 of the main text).

In the opposite limit, $\Omega/\gamma \ll 1$, we use $\mu/\gamma = i \frac{1}{4}+ \mathcal{O}((\Omega/\gamma)^2)$. This yields
\begin{equation}
   \Bigl({\rm Tr}\left(\sigma_z e^
{\ell t} \left(\left|\downarrow\right\rangle \left\langle \downarrow \right | \right) \right) \Bigr)^2 = e^{-\gamma t}[\cosh(\gamma t) +\sinh(\gamma t)],
\end{equation}
and
\begin{equation}
    p_0(t)=e^{-\gamma t/2}[\cosh(\gamma t/2)+\sinh(\gamma t/2)],
    \end{equation}
to leading order in $\Omega/\gamma$. At the same time, from Eq.~\eqref{eq:wL} we see that $\tilde{w}(z)\to 1$ for $z\to 0$, which means that we cannot use the Laplace transform and the geometric series expansion of $[1-\tilde{w}(z)]^{-1}$. Instead, we allow at most one jump in the approach to the steady state ($n=1$) and we directly appeal to the Dyson expansion in the time domain. The asymptotic expression for the variance, then, reads:
\begin{equation}
  \boxed{{\rm Var}(\sigma_z)=e^{-\gamma t/2} \frac{\cosh{\gamma t} +\sinh{\gamma t}}{\cosh{(\gamma t/2)} + \sinh{(\gamma t/2)}}-1 + \mathcal{O}((\Omega/\gamma)^2)=0 + \mathcal{O}((\Omega/\gamma)^2).} 
\end{equation}
In fact, Monte-Carlo simulations show that ${\rm Var}(\sigma_z) \sim 10^{-6}$ for $\gamma/\Omega=56$.


\subsection{The moment-based method and the truncated hierarchy of equations: Poisson- and Wiener-type unraveling}
	There are two principal unravelings which define the evolution of the conditioned state vector $\ket{\psi_c}=\ket{\psi_c(t)}$. One is driven by Wiener noise~\cite{BarBel91, GP92} (in this section we set $\hbar=1$ and denote $\langle A(t) \rangle$ by $\langle A \rangle_t$)
	\begin{align}
		d\ket{\psi_c}=-iH\ket{\psi_c}dt+&\sum_i\left( \expc{L_i^\dagger}_tL_i-\frac{1}{2}L^\dagger_i L_i-\frac{1}{2}|\expc{L_i}_t|^2
		\right)\ket{\psi_c}dt\nonumber\\
		&+\frac{1}{\sqrt{2}}\sum_i \left(L_i-\expc{L_i}_t\right)\ket{\psi_c}dW_i(t),
		\label{GPeqn}
	\end{align}
where $W_i$ is a complex Wiener process with It\^o rule $dW_i^*dW_j=dW_idW_j^*=2\delta_{ij}dt$ and all others zero.  The other unravelling is driven by Poisson noise~\cite{BarBel91},
	\begin{align}
		d\ket{\psi_c}=-\bigg(iH&+\frac{1}{2}\sum_i L_i^\dagger L_i-\expc{L_i^\dagger L_i}_t\bigg)\ket{\psi_c}dt\nonumber\\
		&+\sum_i \left(\dfrac{L_i}{\langle L_i^\dagger L_i\rangle_{t-}^{1/2}}-I\right)\ket{\psi_{t-}}dN_i(t),
		\label{PDP}
	\end{align}
where $N_i$ are real Poisson processes with It\^o rule $dN_idN_j=\delta_{ij}dN_i$ and $dN_idt=0$.

We can derive the evolution of the expectation of an observable $A$ by using the It\^o product formula~\cite{KeysPhD}, $d\expc{A}=d\expc{\psi_c|A|\psi_c}=(d\bra{\psi_c})A\ket{\psi_c}+\bra{\psi_c}A(d\ket{\psi_c})
	+(d\bra{\psi_c})A(d\ket{\psi_c})$.
	The resulting equations are
	\begin{equation}
	d\expc{A}_t=\expc{\mathcal{L}^\dagger[A]}_tdt+\dfrac{1}{\sqrt{2}}\left[\sum_j\expc{A(L_j-\expc{L_j})}_tdW_j(t)+h.c.\right]
	\label{expcwien}
	\end{equation}
	and
	\begin{equation} 
		d\expc{A}_t=\expc{\mathcal{L}^\dagger[A]}_t dt+\sum_j\left(\dfrac{\expc{L_j^\dagger AL_j}_{t-}}{\expc{L_j^\dagger L_j}_{t-}}-\expc{A}_{t-}\right)d\widetilde{N}_j,
	\label{expcpoiss}
	\end{equation}
	where $d\widetilde{N}_j=dN_j-\expc{L_j^\dagger L_j}_t dt$ is the compensated Poisson process (which in the main text assumed the form $d\widetilde{N}=dN-\gamma\expc{\sigma_+\sigma_-}_tdt$).

  Let us denote by $X_i$ the basis operators of the space of observables, with $x_i=\expc{X_i}_t$. In this basis the operators $A$ are described by vectors $a$, with
	 $a_i=\mbox{Tr}[A^\dagger X_i]$ so that expectations become inner products $\expc{A}_t=(a,x)$. Powers of quantum expectations become powers of inner products and thus if
     the ensemble average is then taken of these powers, the problem of describing their evolution reduces to finding the evolution of moments, i.e. terms like $\Expc x_i x_j x_k$. Using equations~\ref{expcwien} and \ref{expcpoiss}, we can write the evolution equations of $x_i$ as
	 \begin{equation}
	 	dx_i = (u^i,x)dt+\sum_j f^i_jdW_j+f^{i*}_jdW_j^*
	 \end{equation}
	 in the Wiener case with $u^i$ the vector corresponding to $\mathcal{L}^{\dagger}[X_i]$ and $f^i_j=(f^i_j)_t=\expc{X_i(L_j-\expc{L_j})}_t$.
	 In the Poisson case we can similarly write
	 \begin{equation}
	 	dx_i=(u^i,x)dt+\sum_j g^i_jd\widetilde{N}_j,
	 \end{equation}
	 with $g^i_j = (g^i_j)_t = \dfrac{1}{\expc{L_j^\dagger L_j}_t}\left(\expc{L_j^\dagger X_iL_j}_t-\expc{L_j^\dagger L_j}_t\expc{X_i}_t\right)$. To calculate the evolution of terms like $\Expc x_ix_jx_k$, we apply  the It\^o product formula iteratively to get $d(x_ix_jx_k)=(dx_i)x_jx_k+x_id(x_jx_k)+dx_id(x_jx_k)$, then take the expectation so that all martingale terms cancel out. In the Poisson case we use the fact that for a Poisson integral
	 \begin{equation}
	 \Expc\int_0^tf(s)dN_j(s)=\Expc\int_0^tf(s)\expc{L_j^\dagger L_j}_sds,
	 \end{equation}
	 which gives us license to replace $dN_j(t)$ with $\expc{L_j^\dagger L_j}_t dt$.
 \begin{figure}
\centering
    \includegraphics[width=1\textwidth]{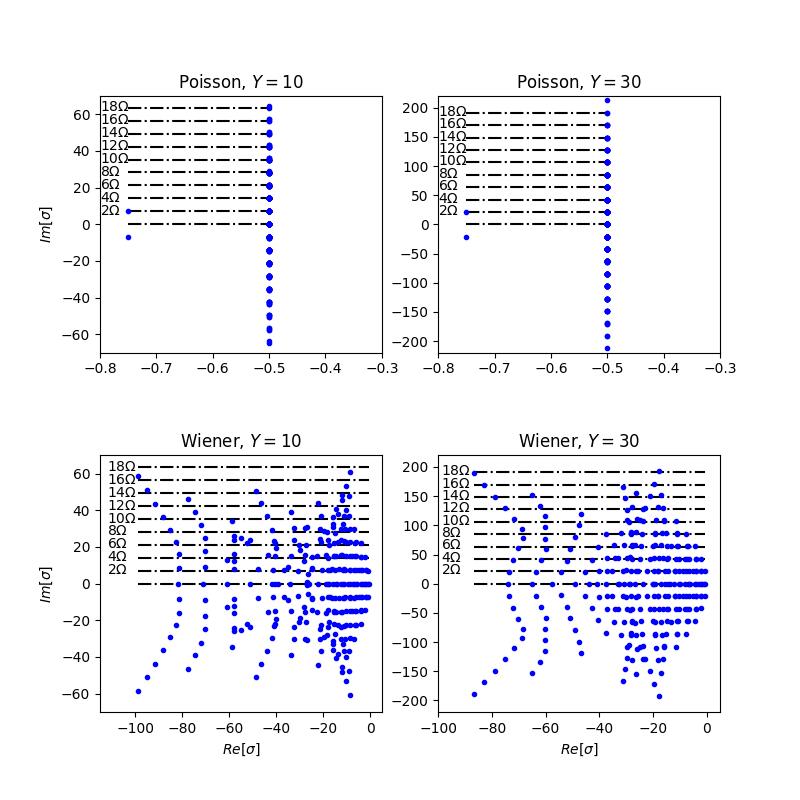}
    \caption{Eigenvalues of the dynamical matrix $M$ corresponding to the Poisson and Wiener-type unravelings for $Y=10, 30$, as indicated on top of each frame. For a Wiener-type unraveling we note that, as $Y$ grows, the spectrum of $M$ aligns better with integer multiples of the Rabi frequency $\Omega_R=2\Omega$.}
        \label{fig:EigFig}
\end{figure}
	 Applying these rules leads to very simple combinatorial expressions for the evolution of moments. In the Wiener case, for the second moment we have
	 \begin{equation}
	 d\Expc x_ix_j=\Expc\left[(u^i,x)x_j+x_i(u^j,x)+\sum_kf^{i*}_kf^j_k+f^i_kf_k^{j*}\right]dt.
	 \end{equation}
	 Due to the structure of the It\^o product formula, simple combinatorial patterns arise for higher moments:
	 \begin{equation}
	\begin{aligned}
		d\Expc x_{i_1}\cdots x_{i_n}&=\Expc\sum_{j=1}^n (u^{i_j},x)x_{i_1}\cdots \hat{x}_{i_j}\cdots x_{i_n}dt+\\
		&\Expc\sum_{k}\sum_{\sigma\in C_2}f^{\sigma(i_1)}_kf^{\sigma(i_2)*}_kx_{\sigma(i_3)}\cdots x_{\sigma(i_n)}dt,
	\end{aligned}
	\end{equation}
	where $C_k$ is the set $\{i_1,\ldots,i_n\}$ choose $k$
	 and hat denotes omission. For the Poisson case the pattern is a bit more complex. If $l_k$ denotes the vector representing $L_k^\dagger L_k$, then we have for the quadratic case
	 \begin{equation}
	d\Expc x_ix_j=\Expc\left[(u^i,x)x_j+x_i(u^j,x)+\sum_kg_k^ig_k^j(l^k,x)\right]dt
	\end{equation}
	and for the general case
	\begin{equation}
	\begin{aligned}
		d\Expc x_{i_1}\cdots x_{i_n}&=\Expc\left[\sum_{j=1}^n(u^{i_j},x)x_{i_1}\cdots \hat{x}_{i_j}\cdots x_{i_n}\right.
		+\sum_k\sum_{\sigma\in C_2}g_k^{\sigma(i_1)}g_k^{\sigma(i_2)}x_{\sigma(i_3)}\cdots x_{\sigma(i_n)}(l^k,x)\\
		&\left.+\sum_k\sum_{\sigma\in C_3}g_k^{\sigma(i_1)}g_k^{\sigma(i_2)}g_k^{\sigma(i_3)}x_{\sigma(i_4)}\cdots x_{\sigma(i_n)}(l^k,x)+\cdots+\sum_kg_k^{i_1}\cdots g_k^{i_n}(l_k,x)\right]dt,
		\end{aligned}
	\end{equation}
	where in this case there is a summation for every set $C_l$, $l=1,\ldots,n$, each having $l$ many $g^{\sigma(i)}_k$ terms. 
	\par
\begin{figure}
    \centering
    \includegraphics[width=\textwidth]{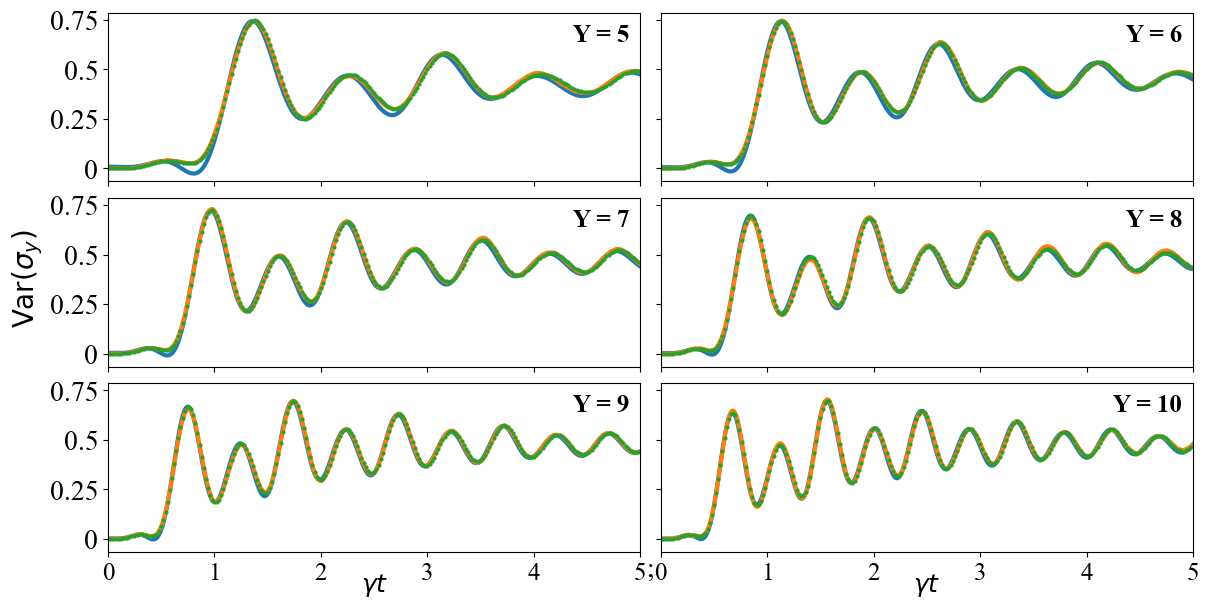}
    \caption{Comparison of ${\rm Var}(\sigma_y)$ plotted against the dimensionless time $\gamma t$ for the ideal case $\overline{n}=0$ and $\eta=1$, with three different methods and for six different driving strengths $Y$, indicated in the top-right corner of each panel. The Monte-Carlo average over $10^4$ realizations is plotted in orange. The blue line and green dots correspond to the analytical results obtained from the Dyson-series expansion to first order in $\gamma/\Omega$, and the moment-based equations, respectively. As the driving strength increases, so does the agreement between the three methods. Note the negative artefact in the short-time evolution of ${\rm Var}(\sigma_y)$, which disappears when higher-order terms in $\gamma^2/\Omega^2$ are considered in Eq.~\eqref{eq:Om}. We also remark that in the long-time limit and for $Y \gg 1$, ${\rm Var}(\sigma_y) + {\rm Var}(\sigma_z)=\frac{1}{2} + \frac{1}{2}=1$. As in Fig. 2 of the main text, the atom is initialized to its ground state.}
    \label{fig:SM_Comparison_3_methods}
\end{figure}

	 In both cases, the evolution is described by a linear equation in the moments (even in the Poisson case where a polynomial division must be performed which has remainder $0$), however the equations for the evolution of a moment generally contain higher order moments so the equations do not close. One way to handle this is to truncate at some high order and not consider the evolution of terms beyond that order. Then the system of equations can be solved using standard methods. We can collect the elementary moments $\Expc x_{i_1}\cdots x_{i_n}$ into a vector $y$ and the coefficients of the moments in each equation $d\Expc x_{j_1}\cdots x_{j_n}$ to get a matrix, $M$. This matrix can be exponentiated to arrive at a time evolution for the vector $y$.  To obtain the spectrum of $M$ and the results depicted in Fig. 2 of the main text, a 10$^{\rm th}$ order truncation of the iterative scheme was used. Eigenvalues of the matrix $M$ for $Y=10,30$ are plotted in Fig.~\ref{fig:EigFig}. They show a banded structure at even multiplies of $\Omega$, which gives rise to the characteristic periodicity in the solutions of ${\rm Var}(\sigma_z)$. For homodyne and heterodyne detection, closely-spaced eigenvalues of the same imaginary parts give rise to ``destructive interference'', dephasing the variance, as we have seen in Fig. 2 of the main text. A similar dephasing is noted in the QTAV ${\rm Var}(\sigma_y)$ for direct phototodetection, shown in Fig.~\ref{fig:SM_Comparison_3_methods}. Once more, we note the good agreement between the Monte-Carlo simulations, the moment-based method and the Dyson-expansion perturbative treatment to first order in $\gamma/\Omega$. Evidently, the agreement betters for increasing $Y$.  

 \subsection{Direct photoelectron-counting with imperfect detectors and/or a thermal bath}

For a limited detector efficiency $\eta<1$ we cannot use a pure state to describe the evolution between collapses. Instead, the general form must be implemented in density matrix form. Between collapses the propagation rule reads
\begin{equation}\label{eq:coheta}
    \ell^{\prime} \rho_c=(\mathcal{L}-\eta J)\rho_c=\frac{1}{i\hbar}[H, \rho_c] + \gamma(1-\eta) \sigma_- \rho_c \sigma_+ - \frac{\gamma}{2}\{\rho_c, \sigma_+ \sigma_-\},
\end{equation}
($\{.,.\}$ stands for the anti-commutator) while the collapse probability for the interval $(t,t+\Delta t]$ is
\begin{equation}
 p_c(t)=\eta {\rm tr}[J \rho_c(t)]\Delta t =\eta(\gamma\Delta t) {\rm tr}[\rho_c(t)\sigma_+ \sigma_-],
\end{equation}
and the (un-normalized) state becomes $\eta J \rho_c$. For $\eta\ll 1$, a single trajectory closely follows the deterministic evolution governed by the ME. This explains the significant reduction in the variance (the case for very weak drive), which -- as we have seen in the main text -- ``responds" to quantum jumps and the oscillatory regression of the fluctuation. 

If we now admit a thermal light injecting a photon flux $\gamma \overline{n}$, then the following term is added to the coherent evolution between collapses [to RHS of Eq.~\eqref{eq:coheta}]:
\begin{equation}
    \gamma \overline{n}(\sigma_- \rho_c \sigma_+ + \sigma_+ \rho_c \sigma_- - \sigma_+\sigma_-\rho_c - \rho_c \sigma_- \sigma_+). 
\end{equation}
In the ideal case, $\overline{n}=0,\,\eta=1$, the series of photon ``clicks'' fully defines the quantum trajectory, since the wavefunction evolves under the action of $H_{\rm eff}$ defined in Eq.~\eqref{eq:Heff}, being reset to the ground state after a spontaneous emission occurs. At optical frequencies, one has $\overline{n}\ll 1$, whence the most detrimental factor to the coherence of individual realizations is the limited detector efficiency. In that case, the propagation rule of Eq.~\eqref{eq:coheta} must be used between jumps which, for $\eta\ \ll 1$, coincides with the action of the Lindblad superoperator $\mathcal{L}$. Photoelectron counting and waiting-time distributions for nonunit detection efficiency are presented in~\cite{Carmichael1989}.

\subsection{Homodyne detection and quadrature amplitude squeezing}

Let us briefly discuss a third type of unravelling (one of Wiener type), in addition to direct photodetection and heterodyne detection, the latter being equivalent to the quantum-state diffusion model~\cite{GP92}. In 1981, Walls and Zoller~\cite{WallsZoller1981} reported that light scattered in resonance fluorescence is squeezed in the field quadrature that is in phase with the mean scattered field amplitude, proportional to $\braket{\sigma_-}_{\rm ss}=+iY/(1+Y^2)$ in the steady state, whence in a direction along $\theta=\pi/2$. A year later, Mandel came up with a scheme for detecting squeezing, which involved homodyning the scattered light with a strong local oscillator and measuring photon counting statistics as a function of the local oscillator phase~\cite{Mandel1982}. Following this approach, the (un-normalized) conditional wavefunction evolves according to the stochastic Schr\"{o}dinger equation
\begin{equation}
    \frac{d}{dt}\ket{\overline{\psi}_c}=\frac{1}{i\hbar}H_W(t)\ket{\overline{\psi}_c},
\end{equation}
in which $H_W(t)$ is the stochastic non-Hermitian Hamiltonian:
\begin{equation}
    H_W(t)=H-i\hbar\gamma \sigma_+ \sigma_- + i\hbar[\sqrt{\gamma} \braket{\psi_c(t)|(e^{i\theta}\sigma_+ + e^{-i\theta}\sigma_-)|\psi_c(t)} + \eta_W(t)]e^{-i\theta}\sqrt{\gamma}\sigma_-,
\end{equation}
where $\theta$ is the local-oscillator phase, $\ket{\psi_c(t)}$ is the normalized state and $\eta_W$ is a Gaussian white noise. Instances of the conditional variance ${\rm Var}(\sigma_z)$ under this unravelling are depicted in Fig. 2 of the main text for $\theta=\pi/2$ and $\theta=0$. For an imperfect detector, the noise $\eta_W$ is replaced by two uncorrelated noise sources added in the proportion $\eta$ and $1-\eta$, with the former featuring in the photocurrent while the latter not~\cite{carmichael1993open}. In Fig. 2 of the main text (bottom panels) we show that the QTAV ${\rm Var}(\sigma_z)$ captures the redistribution of fluctuations among the squeezed and anti-squeezed quadratures. Finally, we recall that in heterodyne detection, $\theta$ is effectively replaced by $-(\omega_{LO}-\omega_A) t$ (see Fig. 1 of the main text). The frequency mismatch between the local oscillator and the drive (here resonant with the atom) is assumed to be very large in comparison to the field fluctuations ($\sim \gamma$). A time average over the period $2\pi/(\omega_{LO}-\omega_A)$ is then performed to simplify the resulting stochastic Schr\"{o}dinger equation. 

\subsection{EPR steering for system-environment entanglement}

We will now digress a little to discuss the relevance of the QTAV and its evolution vis-a-vis the discussion of Wiseman and Gambetta in Ref.~\cite{Wiseman2012} on EPR steering for resonance fluorescence with $Y \gg 1$. In such a configuration, Bob has direct access to the coherently driven two-state system while Alice is situated in the environment where she is able to alternate between unravelings. Then the objective quantum state assumption entails the inequality~\cite{Jones2007}
\begin{equation}\label{ineq:Sdef}
    S(\rho^{c, D},\rho^{c, H})\equiv \overline{f_1(\rho^{c, D})} + \overline{f_2(\rho^{c, H})} \leq 1,
\end{equation}
where $f_1(\rho)\equiv({\rm tr}[\sigma_x \rho])^2$ and $f_2(\rho)\equiv({\rm tr}[\sigma_y \rho])^2 + ({\rm tr}[\sigma_z \rho])^2$; here $\rho$ a state conditioned on a particular measurement scheme selected by Alice.

In assessing whether the EPR steering inequality can be violated in our case, the superscripts $D$ and $H$ denote direct photodetection in vacuum ($\overline{n}=0$) and heterodyne detection, respectively (the latter performed with unit efficiency). Therefore, $\rho^{c, D}$ solves ME~\eqref{eq:ME} with $\ell \to \ell^{\prime}$ from Eq.~\eqref{eq:coheta}, yielding a mixed state, while $\rho^{c, H}$ solves Eq.~\eqref{GPeqn} yielding a pure state at all times.
\begin{figure}[ht]
    \centering
    \includegraphics[width=\textwidth]{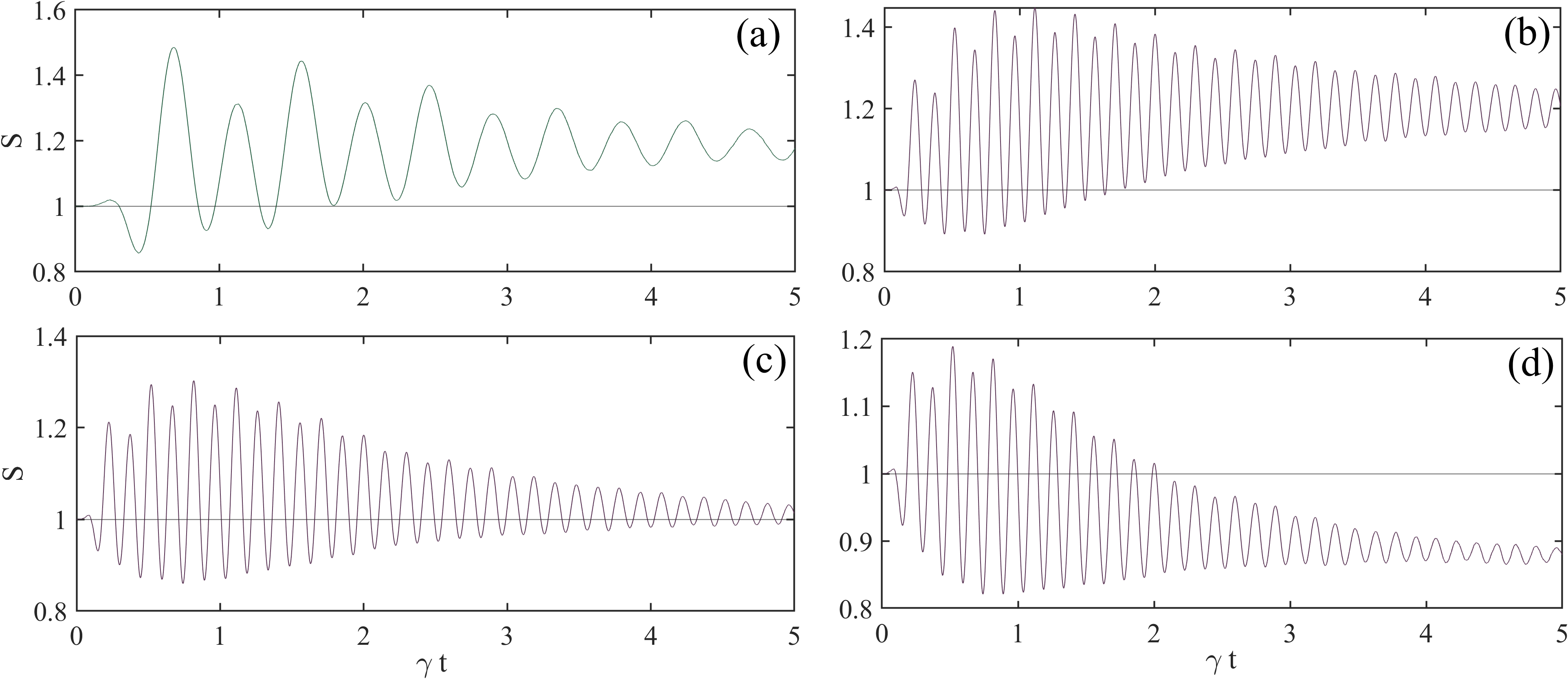}
    \caption{EPR steering $S=S(\rho^{c, D},\rho^{c, H})$ value from the inequality~\eqref{ineq:Sdef} for: $Y=10$ and $\eta=1$ in {\bf (a)}, $Y=30$ and $\eta=1$ in {\bf (b)}, $Y=30$ and $\eta=0.8$ in {\bf (c)}, and $Y=30$ and $\eta=0.6$ in {\bf (d)}.}
    \label{fig:Sv}
\end{figure}

Figure~\ref{fig:Sv} plots the value of $S$ for different values of $Y$ and $\eta$. As noted in~\cite{Wiseman2012}, for $S>1$ the experiment would rule out all theories of objective atomic state reduction. For unit-efficiency  direct detection ($\eta=1$), the steady-state value of $S$ always exceeds unity. We note the envelope is primarily determined by the detection efficiency while the carrier frequency is set by $Y$. Figures~\ref{fig:Sv}(c, d) show that the envelope drops below unity for $\eta \lesssim 0.8$, satisfying the inequality~\eqref{ineq:Sdef}, whence making the experiment ``not provably measurement dependent". In contrast, we note that the feedback scheme adopted for direct photodetection in~\cite{Wiseman2012} guarantees $S>1$ for any value of $\eta$ and a perfect heterodyne detector.


\subsection{Experimental setup and considerations}

In this section, we describe the different experimental imperfections that can affect the measurements of a linear average of $\sigma_z$: the second-order correlation $g^{(2)}(\tau)$. The same spirit could be applied for non-linear averages. Second-order correlation functions are routinely measured experimentally using photon counting techniques in a Hanbury Brown and Twiss configuration. The photon flux enters a beam-splitter and both outputs are monitored by two detectors after a long integration time.

\begin{figure}
    \centering
    \includegraphics[width=1\textwidth]{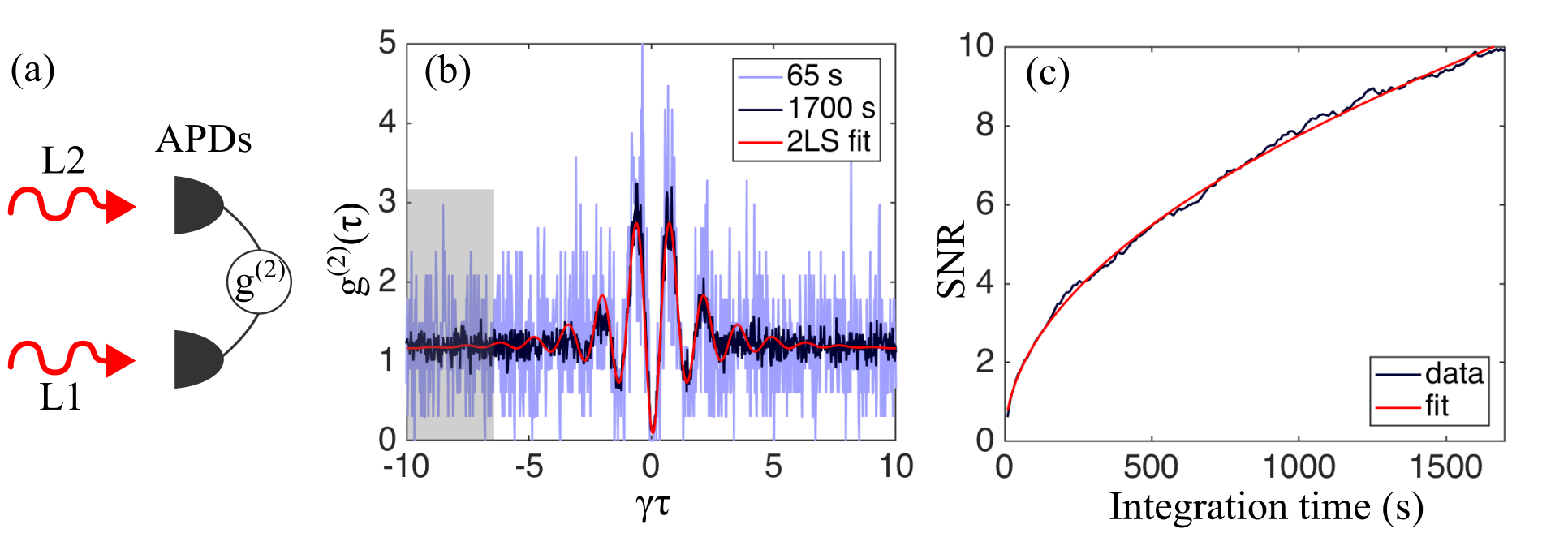}
    \caption{{\bf (a)} The arrival times of the photons coming from the lenses L1 and L2 are measured by avalanche photodiodes (APDs). {\bf (b)} Normalized second-order correlation function for integration times of 65 (blue) or 1700 (black) seconds and fit using Eq.~\eqref{eq:g2_fit_fnc} from a two-level system approximation including experimental imperfections (red). The gray shaded box corresponds to the time range where the signal-to noise of $g^{(2)}$ is computed and shown in {\bf (c)} as a function of the integration time and fitted by Eq.~\eqref{eq:SNR_fit_fnc}.}
    \label{fig:SM_exp}
\end{figure}

We consider real experimental data based on the setup of~\cite{Bruno2019} in order to highlight the different imperfections. The experimental setup consists of a Maltese-cross coupling of single neutral atoms of $^{87}$Rb. The atom is illuminated by a laser field with Rabi frequency $\Omega$ and detuning $\Delta$. The photons are collected by high-numerical aperture lenses, as shown in the main text, and the photon arrival times are measured [Fig. \ref{fig:SM_exp}(a)] using single-photon avalanche photodiodes (APDs). From these measurements, the normalized second-order correlation function between the lenses L1 and L2 is computed [Fig. \ref{fig:SM_exp}(b)]. The average atomic scattering rate on a single detector is $R_{\text{sca,det}}=9000$ counts/s. 

Let us focus on the measured correlation $g^{(2)}_{\text{mes}}(\tau)$ after a long integration time. The fit function for the correlation based on a two-level atom model and including different experimental imperfections is given by:
\begin{equation}
    g^{(2)}_{\text{mes}}(\tau) = \frac{A(\tau)g^{(2)}(\tau,\Delta,\Omega)+\frac{2}{{\rm SNR}_{\text{Det}}}+\frac{1}{{\rm SNR}_{\text{Det}}^2}}{1+\frac{2}{{\rm SNR}_{\text{Det}}}+\frac{1}{{\rm SNR}_{\text{Det}}^2}},
    \label{eq:g2_fit_fnc}
\end{equation}
where ${\rm SNR}_{\text{Det}}=R_{\text{sca,det}}/R_{\text{DC}}=18$ accounts for false positive photon detections due to the dark counts of the detectors. In the experiment, the dark count rate was $R_{\text{DC}}=500$ counts/s for each detector. Then, for single atoms, the photon counts are usually measured using red-detuned light in order to maintain the atom in the trap while acquiring the resonance fluorescence signal. A large negative detuning $\Delta<0$ causes an overshoot of the maximum value of $g^{(2)}$ above $2$. It is taken into account by an explicit dependence on the detuning of $g^{(2)}(\tau,\Delta,\Omega)$ which comes from the steady-state solution of the optical Bloch equations for a two-level atom. Finally, an empirical global envelop $A(\tau)=a+be^{-c\tau}$ takes into account the effect of the atomic motion in the trap~\cite{Markus2006} where the parameters $a, b, c$ are extracted from a fit on microsecond timescales. After including these corrections, the correlation function is fitted from which we extract the Rabi frequency $\Omega=3.3\gamma$ and the detuning $\Delta=-3.2\gamma$.
As shown on Fig.~\ref{fig:SM_exp}(c), modeling the atom as a two-level system including  experimental imperfections is enough to explain the experimental data. This agreement justifies the use of a two-level atom model to compute the different unravelings. 

Now, let us consider that each photon flux arriving on the two detectors is measured with same efficiency $\eta$. In an optical detection system, $\eta$ is given by the product of the collection efficiency $\eta_{\text{col}}$, the optical path losses $\eta_{\text{loss}}$ and the quantum efficiency of the detectors $\eta_{\text{QE}}$. The $g^{(2)}$ statistics is recovered by measuring coincidences on two detectors over a large dataset after a long integration time $t_{\text{int}}$. In the long time limit ($\gamma\tau\gg1$), where the counts are uncorrelated, the coincidences follow a Poissonian statistics. Therefore, the signal-to-noise ${\rm SNR}$ of the measured $g^{(2)}$ scales as: 
\begin{equation}
    {\rm SNR} = \sqrt{\eta^2g^{(2)}R_{\text{sca,det}}t_{\text{int}}}.
    \label{eq:SNR_fit_fnc}
\end{equation}
Experimentally, we evaluate the signal-to-noise of the measured $g^{(2)}_{mes}$ [Fig. \ref{fig:SM_exp}(c)] by computing the ratio of the average and the standard deviation in the steady-state limit [gray shaded area in Fig. \ref{fig:SM_exp}(b)].

By fitting the signal-to-noise using Eq. \eqref{eq:SNR_fit_fnc}, we end up with $\eta=0.25\%$. First, the optical losses $\eta_{\text{loss}}=50\%$ can be improved by a better mode-matching of the emitting spatial modes and the collecting optical fibers. Second, the detector quantum efficiency $\eta_{\text{QE}}=50\%$ at a specific wavelength $\lambda=780$ nm can be increased up to $90\%$ using superconducting nanowires~\cite{Natarajan_2012}. Finally, for an atom scattering photons isotropically, the collection efficiency $\eta_{\text{col}}$ is determined by the solid angle covered by the optical system. Based on these numbers, we deduce the experimental collection efficiency of a single lens of $1\%$. By summing all channels up, the Maltese-cross coupling scheme employing the four lenses increases in principle this coupling by a factor of 4 for an atom tightly trapped in all spatial directions. Other geometries could also be used to increase the collection efficiency, such as a parabolic mirror trap~\cite{Chou2017}.

\vspace{3mm}

\par\noindent\rule{\textwidth}{0.5pt}

\vspace{5mm}

\twocolumngrid

\begin{acknowledgments}
We gratefully acknowledge funding from: ERC AdG NOQIA; MCIN/AEI (PGC2018-0910.13039/501100011033, CEX2019-000910-S/10.13039/501100011033, Plan National FIDEUA PID2019-106901GB-I00, Plan National STAMEENA PID2022-139099NB-I00 project funded by MCIN/AEI/10.13039/501100011033 and by the “European Union NextGenerationEU/PRTR" (PRTR-C17.I1), FPI); QUANTERA MAQS PCI2019-111828-2); QUANTERA DYNAMITE PCI2022-132919 (QuantERA II Programme co-funded by European Union’s Horizon 2020 program under Grant Agreement No 101017733), Ministry of Economic Affairs and Digital Transformation of the Spanish Government through the QUANTUM ENIA project call – Quantum Spain project, and by the European Union through the Recovery, Transformation, and Resilience Plan – NextGenerationEU within the framework of the Digital Spain 2026 Agenda; Fundació Cellex; Fundació Mir-Puig; Generalitat de Catalunya (European Social Fund FEDER and CERCA program, AGAUR Grant No. 2021 SGR 01452, QuantumCAT/U16-011424, co-funded by ERDF Operational Program of Catalonia 2014-2020); Barcelona Supercomputing Center MareNostrum (FI-2023-1-0013); EU Quantum Flagship (PASQuanS2.1, 101113690); EU Horizon 2020 FET-OPEN OPTOlogic (Grant No 899794); EU Horizon Europe Program (Grant Agreement 101080086 — NeQST), ICFO Internal “QuantumGaudi” project; European Union’s Horizon 2020 program under the Marie Sklodowska-Curie grant agreement No 847648; “La Caixa” Junior Leaders fellowships, La Caixa” Foundation (ID 100010434): CF/BQ/PR23/11980043, LCF/BQ/PR23/11980044; European Commission project QUANTIFY (Grant Agreement No. 101135931); Spanish Ministry of Science MCIN: project SAPONARIA (PID2021-123813NB-I00) and ``Severo Ochoa'' Center of Excellence CEX2019-000910-S; Departament de Recerca i Universitats de la Generalitat de Catalunya grant No. 2021 SGR 01453. Views and opinions expressed are, however, those of the author(s) only and do not necessarily reflect those of the European Union, European Commission, European Climate, Infrastructure and Environment Executive Agency (CINEA), or any other granting authority. Neither the European Union nor any granting authority can be held responsible for them. MAG-M acknowledges funding from QuantERA II Cofund 2021 PCI2022-133004, Projects of MCIN with funding from European Union NextGenerationEU (PRTR-C17.I1) and by Generalitat Valenciana, with Ref. 20220883 (PerovsQuTe) and COMCUANTICA/007 (QuanTwin), and Red Tematica RED2022-134391-T. JW was partially supported by NSF grant DMS 1911358 and by the Simons Foundation Fellowship 823539.
\end{acknowledgments}

\clearpage

\bibliography{bibliography}

\begin{thebibliography}{100}%
\makeatletter
\providecommand \@ifxundefined [1]{%
 \@ifx{#1\undefined}
}%
\providecommand \@ifnum [1]{%
 \ifnum #1\expandafter \@firstoftwo
 \else \expandafter \@secondoftwo
 \fi
}%
\providecommand \@ifx [1]{%
 \ifx #1\expandafter \@firstoftwo
 \else \expandafter \@secondoftwo
 \fi
}%
\providecommand \natexlab [1]{#1}%
\providecommand \enquote  [1]{``#1''}%
\providecommand \bibnamefont  [1]{#1}%
\providecommand \bibfnamefont [1]{#1}%
\providecommand \citenamefont [1]{#1}%
\providecommand \href@noop [0]{\@secondoftwo}%
\providecommand \href [0]{\begingroup \@sanitize@url \@href}%
\providecommand \@href[1]{\@@startlink{#1}\@@href}%
\providecommand \@@href[1]{\endgroup#1\@@endlink}%
\providecommand \@sanitize@url [0]{\catcode `\\12\catcode `\$12\catcode
  `\&12\catcode `\#12\catcode `\^12\catcode `\_12\catcode `\%12\relax}%
\providecommand \@@startlink[1]{}%
\providecommand \@@endlink[0]{}%
\providecommand \url  [0]{\begingroup\@sanitize@url \@url }%
\providecommand \@url [1]{\endgroup\@href {#1}{\urlprefix }}%
\providecommand \urlprefix  [0]{URL }%
\providecommand \Eprint [0]{\href }%
\providecommand \doibase [0]{http://dx.doi.org/}%
\providecommand \selectlanguage [0]{\@gobble}%
\providecommand \bibinfo  [0]{\@secondoftwo}%
\providecommand \bibfield  [0]{\@secondoftwo}%
\providecommand \translation [1]{[#1]}%
\providecommand \BibitemOpen [0]{}%
\providecommand \bibitemStop [0]{}%
\providecommand \bibitemNoStop [0]{.\EOS\space}%
\providecommand \EOS [0]{\spacefactor3000\relax}%
\providecommand \BibitemShut  [1]{\csname bibitem#1\endcsname}%
\let\auto@bib@innerbib\@empty
\bibitem [{\citenamefont {Gorini}\ \emph {et~al.}(1976)\citenamefont {Gorini},
  \citenamefont {Kossakowski},\ and\ \citenamefont {Sudarshan}}]{GKS76}%
  \BibitemOpen
  \bibfield  {author} {\bibinfo {author} {\bibfnamefont {Vittorio}\
  \bibnamefont {Gorini}}, \bibinfo {author} {\bibfnamefont {Andrzej}\
  \bibnamefont {Kossakowski}}, \ and\ \bibinfo {author} {\bibfnamefont
  {E.~C.~G.}\ \bibnamefont {Sudarshan}},\ }\bibfield  {title} {\enquote
  {\bibinfo {title} {Completely positive dynamical semigroups of n‐level
  systems},}\ }\href {\doibase 10.1063/1.522979} {\bibfield  {journal}
  {\bibinfo  {journal} {Journal of Mathematical Physics}\ }\textbf {\bibinfo
  {volume} {17}},\ \bibinfo {pages} {821--825} (\bibinfo {year}
  {1976})}\BibitemShut {NoStop}%
\bibitem [{\citenamefont {Lindblad}(1976)}]{Lind76}%
  \BibitemOpen
  \bibfield  {author} {\bibinfo {author} {\bibfnamefont {G.}~\bibnamefont
  {Lindblad}},\ }\bibfield  {title} {\enquote {\bibinfo {title} {On the
  generators of quantum dynamical semigroups},}\ }\href {\doibase
  10.1007/BF01608499} {\bibfield  {journal} {\bibinfo  {journal}
  {Communications in Mathematical Physics}\ }\textbf {\bibinfo {volume} {48}},\
  \bibinfo {pages} {119--130} (\bibinfo {year} {1976})}\BibitemShut {NoStop}%
\bibitem [{\citenamefont {Breuer}\ and\ \citenamefont
  {Petruccione}(1999)}]{BP99}%
  \BibitemOpen
  \bibfield  {author} {\bibinfo {author} {\bibfnamefont {Heinz-Peter}\
  \bibnamefont {Breuer}}\ and\ \bibinfo {author} {\bibfnamefont {Francesco}\
  \bibnamefont {Petruccione}},\ }\bibfield  {title} {\enquote {\bibinfo {title}
  {Stochastic unraveling of relativistic quantum measurements},}\ }in\
  \href@noop {} {\emph {\bibinfo {booktitle} {Open Systems and Measurement in
  Relativistic Quantum Theory}}},\ \bibinfo {editor} {edited by\ \bibinfo
  {editor} {\bibfnamefont {Heinz-Peter}\ \bibnamefont {Breuer}}\ and\ \bibinfo
  {editor} {\bibfnamefont {Francesco}\ \bibnamefont {Petruccione}}}\ (\bibinfo
  {publisher} {Springer Berlin Heidelberg},\ \bibinfo {address} {Berlin,
  Heidelberg},\ \bibinfo {year} {1999})\ pp.\ \bibinfo {pages}
  {81--116}\BibitemShut {NoStop}%
\bibitem [{\citenamefont {Haroche}\ and\ \citenamefont
  {Raimond}(2006)}]{HarocheBook}%
  \BibitemOpen
  \bibfield  {author} {\bibinfo {author} {\bibfnamefont {Serge}\ \bibnamefont
  {Haroche}}\ and\ \bibinfo {author} {\bibfnamefont {Jean-Michel}\ \bibnamefont
  {Raimond}},\ }\href {\doibase 10.1093/acprof:oso/9780198509141.001.0001}
  {\emph {\bibinfo {title} {{Exploring the Quantum: Atoms, Cavities, and
  Photons}}}}\ (\bibinfo  {publisher} {Oxford University Press},\ \bibinfo
  {year} {2006})\BibitemShut {NoStop}%
\bibitem [{\citenamefont {Breuer}\ and\ \citenamefont
  {Petruccione}(2002)}]{BP02}%
  \BibitemOpen
  \bibfield  {author} {\bibinfo {author} {\bibfnamefont {Heinz~Peter}\
  \bibnamefont {Breuer}}\ and\ \bibinfo {author} {\bibfnamefont {Francesco}\
  \bibnamefont {Petruccione}},\ }\href@noop {} {\emph {\bibinfo {title} {The
  theory of open quantum systems}}}\ (\bibinfo  {publisher} {Oxford university
  press},\ \bibinfo {year} {2002})\BibitemShut {NoStop}%
\bibitem [{\citenamefont {Attal}\ \emph {et~al.}(2006)\citenamefont {Attal},
  \citenamefont {Joye},\ and\ \citenamefont {Pillet}}]{OQS206}%
  \BibitemOpen
  \bibinfo {editor} {\bibfnamefont {S.}~\bibnamefont {Attal}}, \bibinfo
  {editor} {\bibfnamefont {A.}~\bibnamefont {Joye}}, \ and\ \bibinfo {editor}
  {\bibfnamefont {C.-A.}\ \bibnamefont {Pillet}},\ eds.,\ \href@noop {} {\emph
  {\bibinfo {title} {Open Quantum Systems II -- The Markovian Approach}}},\
  Vol.\ \bibinfo {volume} {1881}\ (\bibinfo  {publisher} {Springer Berlin,
  Heidelberg},\ \bibinfo {year} {2006})\ \bibinfo {note} {lecture Notes in
  Mathematics}\BibitemShut {NoStop}%
\bibitem [{\citenamefont {Carmichael}(2008)}]{CarmichaelBook2}%
  \BibitemOpen
  \bibfield  {author} {\bibinfo {author} {\bibfnamefont {Howard}\ \bibnamefont
  {Carmichael}},\ }\href@noop {} {\emph {\bibinfo {title} {Statistical Methods
  in Quantum Optics 2}}}\ (\bibinfo  {publisher} {Springer, Berlin, Germany},\
  \bibinfo {year} {2008})\ Chap.\ \bibinfo {chapter} {9, 17}\BibitemShut
  {NoStop}%
\bibitem [{\citenamefont {Accardi}\ \emph {et~al.}(2013)\citenamefont
  {Accardi}, \citenamefont {Volovich},\ and\ \citenamefont {Lu}}]{Accardi13}%
  \BibitemOpen
  \bibfield  {author} {\bibinfo {author} {\bibfnamefont {L.}~\bibnamefont
  {Accardi}}, \bibinfo {author} {\bibfnamefont {I.}~\bibnamefont {Volovich}}, \
  and\ \bibinfo {author} {\bibfnamefont {Y.~G.}\ \bibnamefont {Lu}},\
  }\href@noop {} {\emph {\bibinfo {title} {Quantum Theory and Its Stochastic
  Limit}}}\ (\bibinfo  {publisher} {Springer Berlin, Heidelberg},\ \bibinfo
  {year} {2013})\BibitemShut {NoStop}%
\bibitem [{\citenamefont {N.~Bohr}\ and\ \citenamefont
  {Slater}(1924)}]{BKS1924}%
  \BibitemOpen
  \bibfield  {author} {\bibinfo {author} {\bibfnamefont {H.A.~Kramers}\
  \bibnamefont {N.~Bohr}}\ and\ \bibinfo {author} {\bibfnamefont {J.C.}\
  \bibnamefont {Slater}},\ }\bibfield  {title} {\enquote {\bibinfo {title}
  {Lxxvi. the quantum theory of radiation},}\ }\href {\doibase
  10.1080/14786442408565262} {\bibfield  {journal} {\bibinfo  {journal} {The
  London, Edinburgh, and Dublin Philosophical Magazine and Journal of Science}\
  }\textbf {\bibinfo {volume} {47}},\ \bibinfo {pages} {785--802} (\bibinfo
  {year} {1924})}\BibitemShut {NoStop}%
\bibitem [{\citenamefont {Carmichael}(1993)}]{carmichael1993open}%
  \BibitemOpen
  \bibfield  {author} {\bibinfo {author} {\bibfnamefont {Howard}\ \bibnamefont
  {Carmichael}},\ }\href {\doibase 10.1007/978-3-540-47620-7} {\emph {\bibinfo
  {title} {An open systems approach to quantum optics}}}\ (\bibinfo
  {publisher} {Springer, Berlin, Germany},\ \bibinfo {year} {1993})\BibitemShut
  {NoStop}%
\bibitem [{\citenamefont {Gardiner}\ \emph {et~al.}(1992)\citenamefont
  {Gardiner}, \citenamefont {Parkins},\ and\ \citenamefont
  {Zoller}}]{Gardiner92}%
  \BibitemOpen
  \bibfield  {author} {\bibinfo {author} {\bibfnamefont {C.~W.}\ \bibnamefont
  {Gardiner}}, \bibinfo {author} {\bibfnamefont {A.~S.}\ \bibnamefont
  {Parkins}}, \ and\ \bibinfo {author} {\bibfnamefont {P.}~\bibnamefont
  {Zoller}},\ }\bibfield  {title} {\enquote {\bibinfo {title} {Wave-function
  quantum stochastic differential equations and quantum-jump simulation
  methods},}\ }\href {\doibase 10.1103/PhysRevA.46.4363} {\bibfield  {journal}
  {\bibinfo  {journal} {Phys. Rev. A}\ }\textbf {\bibinfo {volume} {46}},\
  \bibinfo {pages} {4363--4381} (\bibinfo {year} {1992})}\BibitemShut {NoStop}%
\bibitem [{\citenamefont {Dalibard}\ \emph {et~al.}(1992)\citenamefont
  {Dalibard}, \citenamefont {Castin},\ and\ \citenamefont {M\o{}lmer}}]{DCM92}%
  \BibitemOpen
  \bibfield  {author} {\bibinfo {author} {\bibfnamefont {Jean}\ \bibnamefont
  {Dalibard}}, \bibinfo {author} {\bibfnamefont {Yvan}\ \bibnamefont {Castin}},
  \ and\ \bibinfo {author} {\bibfnamefont {Klaus}\ \bibnamefont {M\o{}lmer}},\
  }\bibfield  {title} {\enquote {\bibinfo {title} {Wave-function approach to
  dissipative processes in quantum optics},}\ }\href {\doibase
  10.1103/PhysRevLett.68.580} {\bibfield  {journal} {\bibinfo  {journal} {Phys.
  Rev. Lett.}\ }\textbf {\bibinfo {volume} {68}},\ \bibinfo {pages} {580--583}
  (\bibinfo {year} {1992})}\BibitemShut {NoStop}%
\bibitem [{\citenamefont {Omn\`es}(1994)}]{omnes1994}%
  \BibitemOpen
  \bibfield  {author} {\bibinfo {author} {\bibfnamefont {Roland}\ \bibnamefont
  {Omn\`es}},\ }\href@noop {} {\emph {\bibinfo {title} {The Interpretation of
  Quantum Mechanics}}}\ (\bibinfo  {publisher} {Princeton University Press,
  Princeton, New Jersey},\ \bibinfo {year} {1994})\BibitemShut {NoStop}%
\bibitem [{\citenamefont {Blanchard}\ and\ \citenamefont
  {Jadczyk}(1995)}]{Blanchard1995}%
  \BibitemOpen
  \bibfield  {author} {\bibinfo {author} {\bibfnamefont {Ph.}\ \bibnamefont
  {Blanchard}}\ and\ \bibinfo {author} {\bibfnamefont {A.}~\bibnamefont
  {Jadczyk}},\ }\bibfield  {title} {\enquote {\bibinfo {title} {Event-enhanced
  quantum theory and piecewise deterministic dynamics},}\ }\href {\doibase
  https://doi.org/10.1002/andp.19955070605} {\bibfield  {journal} {\bibinfo
  {journal} {Annalen der Physik}\ }\textbf {\bibinfo {volume} {507}},\ \bibinfo
  {pages} {583--599} (\bibinfo {year} {1995})}\BibitemShut {NoStop}%
\bibitem [{\citenamefont {Korotkov}(1999)}]{Korotkov99}%
  \BibitemOpen
  \bibfield  {author} {\bibinfo {author} {\bibfnamefont {Alexander~N.}\
  \bibnamefont {Korotkov}},\ }\bibfield  {title} {\enquote {\bibinfo {title}
  {Continuous quantum measurement of a double dot},}\ }\href {\doibase
  10.1103/PhysRevB.60.5737} {\bibfield  {journal} {\bibinfo  {journal} {Phys.
  Rev. B}\ }\textbf {\bibinfo {volume} {60}},\ \bibinfo {pages} {5737--5742}
  (\bibinfo {year} {1999})}\BibitemShut {NoStop}%
\bibitem [{\citenamefont {Plenio}\ and\ \citenamefont
  {Knight}(1998)}]{Plenio1998}%
  \BibitemOpen
  \bibfield  {author} {\bibinfo {author} {\bibfnamefont {M.~B.}\ \bibnamefont
  {Plenio}}\ and\ \bibinfo {author} {\bibfnamefont {P.~L.}\ \bibnamefont
  {Knight}},\ }\bibfield  {title} {\enquote {\bibinfo {title} {The quantum-jump
  approach to dissipative dynamics in quantum optics},}\ }\href {\doibase
  10.1103/RevModPhys.70.101} {\bibfield  {journal} {\bibinfo  {journal} {Rev.
  Mod. Phys.}\ }\textbf {\bibinfo {volume} {70}},\ \bibinfo {pages} {101--144}
  (\bibinfo {year} {1998})}\BibitemShut {NoStop}%
\bibitem [{\citenamefont {Hegerfeldt}\ and\ \citenamefont
  {Sondermann}(1996)}]{Hegerfeldt1996}%
  \BibitemOpen
  \bibfield  {author} {\bibinfo {author} {\bibfnamefont {Gerhard~C}\
  \bibnamefont {Hegerfeldt}}\ and\ \bibinfo {author} {\bibfnamefont {Dirk~G}\
  \bibnamefont {Sondermann}},\ }\bibfield  {title} {\enquote {\bibinfo {title}
  {Conditional hamiltonian and reset operator in the quantum jump approach},}\
  }\href {\doibase 10.1088/1355-5111/8/1/010} {\bibfield  {journal} {\bibinfo
  {journal} {Quantum and Semiclassical Optics: Journal of the European Optical
  Society Part B}\ }\textbf {\bibinfo {volume} {8}},\ \bibinfo {pages} {121}
  (\bibinfo {year} {1996})}\BibitemShut {NoStop}%
\bibitem [{\citenamefont {Daley}(2014)}]{Daley2014}%
  \BibitemOpen
  \bibfield  {author} {\bibinfo {author} {\bibfnamefont {Andrew~J.}\
  \bibnamefont {Daley}},\ }\bibfield  {title} {\enquote {\bibinfo {title}
  {Quantum trajectories and open many-body quantum systems},}\ }\href {\doibase
  10.1080/00018732.2014.933502} {\bibfield  {journal} {\bibinfo  {journal}
  {Advances in Physics}\ }\textbf {\bibinfo {volume} {63}},\ \bibinfo {pages}
  {77--149} (\bibinfo {year} {2014})}\BibitemShut {NoStop}%
\bibitem [{\citenamefont {Wineland}\ and\ \citenamefont
  {Dehmelt}(1975)}]{Dehmelt}%
  \BibitemOpen
  \bibfield  {author} {\bibinfo {author} {\bibfnamefont {D.~J.}\ \bibnamefont
  {Wineland}}\ and\ \bibinfo {author} {\bibfnamefont {H.}~\bibnamefont
  {Dehmelt}},\ }\href@noop {} {\bibfield  {journal} {\bibinfo  {journal} {Bull.
  Am. Phys. Soc.}\ }\textbf {\bibinfo {volume} {20}},\ \bibinfo {pages} {637}
  (\bibinfo {year} {1975})}\BibitemShut {NoStop}%
\bibitem [{\citenamefont {Dehmelt}(1981)}]{Dehmelt1}%
  \BibitemOpen
  \bibfield  {author} {\bibinfo {author} {\bibfnamefont {H.}~\bibnamefont
  {Dehmelt}},\ }\bibfield  {title} {\enquote {\bibinfo {title} {Coherent
  spectroscopy on a single atomic system at rest in free space ii},}\
  }\href@noop {} {\bibfield  {journal} {\bibinfo  {journal} {J. Phys.
  Colloques}\ }\textbf {\bibinfo {volume} {42}},\ \bibinfo {pages} {C8--299}
  (\bibinfo {year} {1981})}\BibitemShut {NoStop}%
\bibitem [{\citenamefont {Nagourney}\ \emph {et~al.}(1986)\citenamefont
  {Nagourney}, \citenamefont {Sandberg},\ and\ \citenamefont
  {Dehmelt}}]{Nagourney86}%
  \BibitemOpen
  \bibfield  {author} {\bibinfo {author} {\bibfnamefont {Warren}\ \bibnamefont
  {Nagourney}}, \bibinfo {author} {\bibfnamefont {Jon}\ \bibnamefont
  {Sandberg}}, \ and\ \bibinfo {author} {\bibfnamefont {Hans}\ \bibnamefont
  {Dehmelt}},\ }\bibfield  {title} {\enquote {\bibinfo {title} {Shelved optical
  electron amplifier: Observation of quantum jumps},}\ }\href {\doibase
  10.1103/PhysRevLett.56.2797} {\bibfield  {journal} {\bibinfo  {journal}
  {Phys. Rev. Lett.}\ }\textbf {\bibinfo {volume} {56}},\ \bibinfo {pages}
  {2797--2799} (\bibinfo {year} {1986})}\BibitemShut {NoStop}%
\bibitem [{\citenamefont {Sauter}\ \emph {et~al.}(1986)\citenamefont {Sauter},
  \citenamefont {Neuhauser}, \citenamefont {Blatt},\ and\ \citenamefont
  {Toschek}}]{Sauter86}%
  \BibitemOpen
  \bibfield  {author} {\bibinfo {author} {\bibfnamefont {Th.}\ \bibnamefont
  {Sauter}}, \bibinfo {author} {\bibfnamefont {W.}~\bibnamefont {Neuhauser}},
  \bibinfo {author} {\bibfnamefont {R.}~\bibnamefont {Blatt}}, \ and\ \bibinfo
  {author} {\bibfnamefont {P.~E.}\ \bibnamefont {Toschek}},\ }\bibfield
  {title} {\enquote {\bibinfo {title} {Observation of quantum jumps},}\ }\href
  {\doibase 10.1103/PhysRevLett.57.1696} {\bibfield  {journal} {\bibinfo
  {journal} {Phys. Rev. Lett.}\ }\textbf {\bibinfo {volume} {57}},\ \bibinfo
  {pages} {1696--1698} (\bibinfo {year} {1986})}\BibitemShut {NoStop}%
\bibitem [{\citenamefont {Bergquist}\ \emph {et~al.}(1986)\citenamefont
  {Bergquist}, \citenamefont {Hulet}, \citenamefont {Itano},\ and\
  \citenamefont {Wineland}}]{QuantumJumps}%
  \BibitemOpen
  \bibfield  {author} {\bibinfo {author} {\bibfnamefont {J.~C.}\ \bibnamefont
  {Bergquist}}, \bibinfo {author} {\bibfnamefont {Randall~G.}\ \bibnamefont
  {Hulet}}, \bibinfo {author} {\bibfnamefont {Wayne~M.}\ \bibnamefont {Itano}},
  \ and\ \bibinfo {author} {\bibfnamefont {D.~J.}\ \bibnamefont {Wineland}},\
  }\bibfield  {title} {\enquote {\bibinfo {title} {Observation of quantum jumps
  in a single atom},}\ }\href {\doibase 10.1103/PhysRevLett.57.1699} {\bibfield
   {journal} {\bibinfo  {journal} {Phys. Rev. Lett.}\ }\textbf {\bibinfo
  {volume} {57}},\ \bibinfo {pages} {1699--1702} (\bibinfo {year}
  {1986})}\BibitemShut {NoStop}%
\bibitem [{\citenamefont {Basch{\'e}}\ \emph {et~al.}(1995)\citenamefont
  {Basch{\'e}}, \citenamefont {Kummer},\ and\ \citenamefont
  {Br{\"a}uchle}}]{Basche95}%
  \BibitemOpen
  \bibfield  {author} {\bibinfo {author} {\bibfnamefont {Th.}\ \bibnamefont
  {Basch{\'e}}}, \bibinfo {author} {\bibfnamefont {S.}~\bibnamefont {Kummer}},
  \ and\ \bibinfo {author} {\bibfnamefont {C.}~\bibnamefont {Br{\"a}uchle}},\
  }\bibfield  {title} {\enquote {\bibinfo {title} {Direct spectroscopic
  observation of quantum jumps of a single molecule},}\ }\href {\doibase
  10.1038/373132a0} {\bibfield  {journal} {\bibinfo  {journal} {Nature}\
  }\textbf {\bibinfo {volume} {373}},\ \bibinfo {pages} {132--134} (\bibinfo
  {year} {1995})}\BibitemShut {NoStop}%
\bibitem [{\citenamefont {Peil}\ and\ \citenamefont
  {Gabrielse}(1999)}]{Peil99}%
  \BibitemOpen
  \bibfield  {author} {\bibinfo {author} {\bibfnamefont {S.}~\bibnamefont
  {Peil}}\ and\ \bibinfo {author} {\bibfnamefont {G.}~\bibnamefont
  {Gabrielse}},\ }\bibfield  {title} {\enquote {\bibinfo {title} {Observing the
  quantum limit of an electron cyclotron: Qnd measurements of quantum jumps
  between fock states},}\ }\href {\doibase 10.1103/PhysRevLett.83.1287}
  {\bibfield  {journal} {\bibinfo  {journal} {Phys. Rev. Lett.}\ }\textbf
  {\bibinfo {volume} {83}},\ \bibinfo {pages} {1287--1290} (\bibinfo {year}
  {1999})}\BibitemShut {NoStop}%
\bibitem [{\citenamefont {Gleyzes}\ \emph {et~al.}(2007)\citenamefont
  {Gleyzes}, \citenamefont {Kuhr}, \citenamefont {Guerlin}, \citenamefont
  {Bernu}, \citenamefont {Del{\'e}glise}, \citenamefont {Busk~Hoff},
  \citenamefont {Brune}, \citenamefont {Raimond},\ and\ \citenamefont
  {Haroche}}]{Gleyzes07}%
  \BibitemOpen
  \bibfield  {author} {\bibinfo {author} {\bibfnamefont {S{\'e}bastien}\
  \bibnamefont {Gleyzes}}, \bibinfo {author} {\bibfnamefont {Stefan}\
  \bibnamefont {Kuhr}}, \bibinfo {author} {\bibfnamefont {Christine}\
  \bibnamefont {Guerlin}}, \bibinfo {author} {\bibfnamefont {Julien}\
  \bibnamefont {Bernu}}, \bibinfo {author} {\bibfnamefont {Samuel}\
  \bibnamefont {Del{\'e}glise}}, \bibinfo {author} {\bibfnamefont {Ulrich}\
  \bibnamefont {Busk~Hoff}}, \bibinfo {author} {\bibfnamefont {Michel}\
  \bibnamefont {Brune}}, \bibinfo {author} {\bibfnamefont {Jean-Michel}\
  \bibnamefont {Raimond}}, \ and\ \bibinfo {author} {\bibfnamefont {Serge}\
  \bibnamefont {Haroche}},\ }\bibfield  {title} {\enquote {\bibinfo {title}
  {Quantum jumps of light recording the birth and death of a photon in a
  cavity},}\ }\href {\doibase 10.1038/nature05589} {\bibfield  {journal}
  {\bibinfo  {journal} {Nature}\ }\textbf {\bibinfo {volume} {446}},\ \bibinfo
  {pages} {297--300} (\bibinfo {year} {2007})}\BibitemShut {NoStop}%
\bibitem [{\citenamefont {Neumann}\ \emph {et~al.}(2010)\citenamefont
  {Neumann}, \citenamefont {Beck}, \citenamefont {Steiner}, \citenamefont
  {Rempp}, \citenamefont {Fedder}, \citenamefont {Hemmer}, \citenamefont
  {Wrachtrup},\ and\ \citenamefont {Jelezko}}]{Neumann10}%
  \BibitemOpen
  \bibfield  {author} {\bibinfo {author} {\bibfnamefont {Philipp}\ \bibnamefont
  {Neumann}}, \bibinfo {author} {\bibfnamefont {Johannes}\ \bibnamefont
  {Beck}}, \bibinfo {author} {\bibfnamefont {Matthias}\ \bibnamefont
  {Steiner}}, \bibinfo {author} {\bibfnamefont {Florian}\ \bibnamefont
  {Rempp}}, \bibinfo {author} {\bibfnamefont {Helmut}\ \bibnamefont {Fedder}},
  \bibinfo {author} {\bibfnamefont {Philip~R.}\ \bibnamefont {Hemmer}},
  \bibinfo {author} {\bibfnamefont {Jörg}\ \bibnamefont {Wrachtrup}}, \ and\
  \bibinfo {author} {\bibfnamefont {Fedor}\ \bibnamefont {Jelezko}},\
  }\bibfield  {title} {\enquote {\bibinfo {title} {Single-shot readout of a
  single nuclear spin},}\ }\href {\doibase 10.1126/science.1189075} {\bibfield
  {journal} {\bibinfo  {journal} {Science}\ }\textbf {\bibinfo {volume}
  {329}},\ \bibinfo {pages} {542--544} (\bibinfo {year} {2010})}\BibitemShut
  {NoStop}%
\bibitem [{\citenamefont {Vijay}\ \emph {et~al.}(2011)\citenamefont {Vijay},
  \citenamefont {Slichter},\ and\ \citenamefont {Siddiqi}}]{Vijay11}%
  \BibitemOpen
  \bibfield  {author} {\bibinfo {author} {\bibfnamefont {R.}~\bibnamefont
  {Vijay}}, \bibinfo {author} {\bibfnamefont {D.~H.}\ \bibnamefont {Slichter}},
  \ and\ \bibinfo {author} {\bibfnamefont {I.}~\bibnamefont {Siddiqi}},\
  }\bibfield  {title} {\enquote {\bibinfo {title} {Observation of quantum jumps
  in a superconducting artificial atom},}\ }\href {\doibase
  10.1103/PhysRevLett.106.110502} {\bibfield  {journal} {\bibinfo  {journal}
  {Phys. Rev. Lett.}\ }\textbf {\bibinfo {volume} {106}},\ \bibinfo {pages}
  {110502} (\bibinfo {year} {2011})}\BibitemShut {NoStop}%
\bibitem [{\citenamefont {Minev}\ \emph {et~al.}(2019)\citenamefont {Minev},
  \citenamefont {Mundhada}, \citenamefont {Shankar}, \citenamefont {Reinhold},
  \citenamefont {Guti{\'e}rrez-J{\'a}uregui}, \citenamefont {Schoelkopf},
  \citenamefont {Mirrahimi}, \citenamefont {Carmichael},\ and\ \citenamefont
  {Devoret}}]{Minev2019}%
  \BibitemOpen
  \bibfield  {author} {\bibinfo {author} {\bibfnamefont {Z.~K.}\ \bibnamefont
  {Minev}}, \bibinfo {author} {\bibfnamefont {S.~O.}\ \bibnamefont {Mundhada}},
  \bibinfo {author} {\bibfnamefont {S.}~\bibnamefont {Shankar}}, \bibinfo
  {author} {\bibfnamefont {P.}~\bibnamefont {Reinhold}}, \bibinfo {author}
  {\bibfnamefont {R.}~\bibnamefont {Guti{\'e}rrez-J{\'a}uregui}}, \bibinfo
  {author} {\bibfnamefont {R.~J.}\ \bibnamefont {Schoelkopf}}, \bibinfo
  {author} {\bibfnamefont {M.}~\bibnamefont {Mirrahimi}}, \bibinfo {author}
  {\bibfnamefont {H.~J.}\ \bibnamefont {Carmichael}}, \ and\ \bibinfo {author}
  {\bibfnamefont {M.~H.}\ \bibnamefont {Devoret}},\ }\bibfield  {title}
  {\enquote {\bibinfo {title} {To catch and reverse a quantum jump
  mid-flight},}\ }\href {\doibase 10.1038/s41586-019-1287-z} {\bibfield
  {journal} {\bibinfo  {journal} {Nature}\ }\textbf {\bibinfo {volume} {570}},\
  \bibinfo {pages} {200--204} (\bibinfo {year} {2019})}\BibitemShut {NoStop}%
\bibitem [{\citenamefont {Cook}\ and\ \citenamefont {Kimble}(1985)}]{Cook}%
  \BibitemOpen
  \bibfield  {author} {\bibinfo {author} {\bibfnamefont {Richard~J.}\
  \bibnamefont {Cook}}\ and\ \bibinfo {author} {\bibfnamefont {H.~J.}\
  \bibnamefont {Kimble}},\ }\bibfield  {title} {\enquote {\bibinfo {title}
  {Possibility of direct observation of quantum jumps},}\ }\href {\doibase
  10.1103/PhysRevLett.54.1023} {\bibfield  {journal} {\bibinfo  {journal}
  {Phys. Rev. Lett.}\ }\textbf {\bibinfo {volume} {54}},\ \bibinfo {pages}
  {1023--1026} (\bibinfo {year} {1985})}\BibitemShut {NoStop}%
\bibitem [{\citenamefont {Kimble}\ \emph {et~al.}(1986)\citenamefont {Kimble},
  \citenamefont {Cook},\ and\ \citenamefont {Wells}}]{Kimble}%
  \BibitemOpen
  \bibfield  {author} {\bibinfo {author} {\bibfnamefont {H.~J.}\ \bibnamefont
  {Kimble}}, \bibinfo {author} {\bibfnamefont {Richard~J.}\ \bibnamefont
  {Cook}}, \ and\ \bibinfo {author} {\bibfnamefont {Ann~L.}\ \bibnamefont
  {Wells}},\ }\bibfield  {title} {\enquote {\bibinfo {title} {Intermittent
  atomic fluorescence},}\ }\href {\doibase 10.1103/PhysRevA.34.3190} {\bibfield
   {journal} {\bibinfo  {journal} {Phys. Rev. A}\ }\textbf {\bibinfo {volume}
  {34}},\ \bibinfo {pages} {3190--3195} (\bibinfo {year} {1986})}\BibitemShut
  {NoStop}%
\bibitem [{\citenamefont {Erber}\ and\ \citenamefont
  {Putterman}(1985)}]{Erber}%
  \BibitemOpen
  \bibfield  {author} {\bibinfo {author} {\bibfnamefont {T.}~\bibnamefont
  {Erber}}\ and\ \bibinfo {author} {\bibfnamefont {S.}~\bibnamefont
  {Putterman}},\ }\bibfield  {title} {\enquote {\bibinfo {title} {Randomness in
  quantum mechanics---nature's ultimate cryptogram?}}\ }\href {\doibase
  10.1038/318041a0} {\bibfield  {journal} {\bibinfo  {journal} {Nature}\
  }\textbf {\bibinfo {volume} {318}},\ \bibinfo {pages} {41--43} (\bibinfo
  {year} {1985})}\BibitemShut {NoStop}%
\bibitem [{\citenamefont {Javanainen}(1986)}]{Javanainen}%
  \BibitemOpen
  \bibfield  {author} {\bibinfo {author} {\bibfnamefont {J.}~\bibnamefont
  {Javanainen}},\ }\bibfield  {title} {\enquote {\bibinfo {title} {Possibility
  of quantum jumps in a three-level system},}\ }\href {\doibase
  10.1103/PhysRevA.33.2121} {\bibfield  {journal} {\bibinfo  {journal} {Phys.
  Rev. A}\ }\textbf {\bibinfo {volume} {33}},\ \bibinfo {pages} {2121--2123}
  (\bibinfo {year} {1986})}\BibitemShut {NoStop}%
\bibitem [{\citenamefont {Schenzle}\ \emph {et~al.}(1982)\citenamefont
  {Schenzle}, \citenamefont {DeVoe},\ and\ \citenamefont {Brewer}}]{Schenzle}%
  \BibitemOpen
  \bibfield  {author} {\bibinfo {author} {\bibfnamefont {Axel}\ \bibnamefont
  {Schenzle}}, \bibinfo {author} {\bibfnamefont {Ralph~G.}\ \bibnamefont
  {DeVoe}}, \ and\ \bibinfo {author} {\bibfnamefont {Richard~G.}\ \bibnamefont
  {Brewer}},\ }\bibfield  {title} {\enquote {\bibinfo {title} {Phase-modulation
  laser spectroscopy},}\ }\href {\doibase 10.1103/PhysRevA.25.2606} {\bibfield
  {journal} {\bibinfo  {journal} {Phys. Rev. A}\ }\textbf {\bibinfo {volume}
  {25}},\ \bibinfo {pages} {2606--2621} (\bibinfo {year} {1982})}\BibitemShut
  {NoStop}%
\bibitem [{\citenamefont {Cohen-Tannoudji}\ and\ \citenamefont
  {Dalibard}(1986)}]{Cohen}%
  \BibitemOpen
  \bibfield  {author} {\bibinfo {author} {\bibfnamefont {C.}~\bibnamefont
  {Cohen-Tannoudji}}\ and\ \bibinfo {author} {\bibfnamefont {J.}~\bibnamefont
  {Dalibard}},\ }\bibfield  {title} {\enquote {\bibinfo {title} {Single-atom
  laser spectroscopy. looking for dark periods in fluorescence light},}\ }\href
  {\doibase 10.1209/0295-5075/1/9/004} {\bibfield  {journal} {\bibinfo
  {journal} {Europhysics Letters}\ }\textbf {\bibinfo {volume} {1}},\ \bibinfo
  {pages} {441} (\bibinfo {year} {1986})}\BibitemShut {NoStop}%
\bibitem [{\citenamefont {Cook}(1988)}]{Cook88}%
  \BibitemOpen
  \bibfield  {author} {\bibinfo {author} {\bibfnamefont {Richard~J}\
  \bibnamefont {Cook}},\ }\bibfield  {title} {\enquote {\bibinfo {title} {What
  are quantum jumps?}}\ }\href {\doibase 10.1088/0031-8949/1988/T21/009}
  {\bibfield  {journal} {\bibinfo  {journal} {Physica Scripta}\ }\textbf
  {\bibinfo {volume} {1988}},\ \bibinfo {pages} {49} (\bibinfo {year}
  {1988})}\BibitemShut {NoStop}%
\bibitem [{\citenamefont {Grochmalicki}\ and\ \citenamefont
  {Lewenstein}(1989{\natexlab{a}})}]{Grochmalicki1989}%
  \BibitemOpen
  \bibfield  {author} {\bibinfo {author} {\bibfnamefont {Jan}\ \bibnamefont
  {Grochmalicki}}\ and\ \bibinfo {author} {\bibfnamefont {Maciej}\ \bibnamefont
  {Lewenstein}},\ }\bibfield  {title} {\enquote {\bibinfo {title} {Detection of
  decay processes by means of quantum-jump statistics},}\ }\href {\doibase
  10.1103/PhysRevA.40.2517} {\bibfield  {journal} {\bibinfo  {journal} {Phys.
  Rev. A}\ }\textbf {\bibinfo {volume} {40}},\ \bibinfo {pages} {2517--2528}
  (\bibinfo {year} {1989}{\natexlab{a}})}\BibitemShut {NoStop}%
\bibitem [{\citenamefont {Grochmalicki}\ and\ \citenamefont
  {Lewenstein}(1989{\natexlab{b}})}]{Grochmalicki1989II}%
  \BibitemOpen
  \bibfield  {author} {\bibinfo {author} {\bibfnamefont {Jan}\ \bibnamefont
  {Grochmalicki}}\ and\ \bibinfo {author} {\bibfnamefont {Maciej}\ \bibnamefont
  {Lewenstein}},\ }\bibfield  {title} {\enquote {\bibinfo {title} {Detection of
  cavity fields by means of quantum-jump statistics},}\ }\href {\doibase
  10.1103/PhysRevA.40.2529} {\bibfield  {journal} {\bibinfo  {journal} {Phys.
  Rev. A}\ }\textbf {\bibinfo {volume} {40}},\ \bibinfo {pages} {2529--2533}
  (\bibinfo {year} {1989}{\natexlab{b}})}\BibitemShut {NoStop}%
\bibitem [{\citenamefont {Hegerfeldt}\ and\ \citenamefont
  {Plenio}(1992)}]{Hegerfeldt1992}%
  \BibitemOpen
  \bibfield  {author} {\bibinfo {author} {\bibfnamefont {Gerhard~C.}\
  \bibnamefont {Hegerfeldt}}\ and\ \bibinfo {author} {\bibfnamefont
  {Martin~B.}\ \bibnamefont {Plenio}},\ }\bibfield  {title} {\enquote {\bibinfo
  {title} {Macroscopic dark periods without a metastable state},}\ }\href
  {\doibase 10.1103/PhysRevA.46.373} {\bibfield  {journal} {\bibinfo  {journal}
  {Phys. Rev. A}\ }\textbf {\bibinfo {volume} {46}},\ \bibinfo {pages}
  {373--379} (\bibinfo {year} {1992})}\BibitemShut {NoStop}%
\bibitem [{\citenamefont {Hegerfeldt}\ and\ \citenamefont
  {Plenio}(1993)}]{Hegerfeldt1993}%
  \BibitemOpen
  \bibfield  {author} {\bibinfo {author} {\bibfnamefont {Gerhard~C.}\
  \bibnamefont {Hegerfeldt}}\ and\ \bibinfo {author} {\bibfnamefont
  {Martin~B.}\ \bibnamefont {Plenio}},\ }\bibfield  {title} {\enquote {\bibinfo
  {title} {Coherence with incoherent light: A new type of quantum beat for a
  single atom},}\ }\href {\doibase 10.1103/PhysRevA.47.2186} {\bibfield
  {journal} {\bibinfo  {journal} {Phys. Rev. A}\ }\textbf {\bibinfo {volume}
  {47}},\ \bibinfo {pages} {2186--2190} (\bibinfo {year} {1993})}\BibitemShut
  {NoStop}%
\bibitem [{\citenamefont {Hegerfeldt}(2009)}]{Hegerfeldt2009}%
  \BibitemOpen
  \bibfield  {author} {\bibinfo {author} {\bibfnamefont {Gerhard~C.}\
  \bibnamefont {Hegerfeldt}},\ }\enquote {\bibinfo {title} {The quantum jump
  approach and some of its applications},}\ in\ \href {\doibase
  10.1007/978-3-642-03174-8_6} {\emph {\bibinfo {booktitle} {Time in Quantum
  Mechanics - Vol. 2}}},\ \bibinfo {editor} {edited by\ \bibinfo {editor}
  {\bibfnamefont {Gonzalo}\ \bibnamefont {Muga}}, \bibinfo {editor}
  {\bibfnamefont {Andreas}\ \bibnamefont {Ruschhaupt}}, \ and\ \bibinfo
  {editor} {\bibfnamefont {Adolfo}\ \bibnamefont {del Campo}}}\ (\bibinfo
  {publisher} {Springer Berlin Heidelberg},\ \bibinfo {address} {Berlin,
  Heidelberg},\ \bibinfo {year} {2009})\ pp.\ \bibinfo {pages}
  {127--174}\BibitemShut {NoStop}%
\bibitem [{\citenamefont {Keys}\ and\ \citenamefont {Wehr}(2020)}]{KeysWehr}%
  \BibitemOpen
  \bibfield  {author} {\bibinfo {author} {\bibfnamefont {Dustin}\ \bibnamefont
  {Keys}}\ and\ \bibinfo {author} {\bibfnamefont {Jan}\ \bibnamefont {Wehr}},\
  }\bibfield  {title} {\enquote {\bibinfo {title} {Poisson stochastic master
  equation unravelings and the measurement problem: A quantum stochastic
  calculus perspective},}\ }\href {\doibase 10.1063/1.5133974} {\bibfield
  {journal} {\bibinfo  {journal} {Journal of Mathematical Physics}\ }\textbf
  {\bibinfo {volume} {61}},\ \bibinfo {pages} {032101} (\bibinfo {year}
  {2020})}\BibitemShut {NoStop}%
\bibitem [{\citenamefont {Barchielli}\ and\ \citenamefont
  {Belavkin}(1991)}]{BarBel91}%
  \BibitemOpen
  \bibfield  {author} {\bibinfo {author} {\bibfnamefont {A}~\bibnamefont
  {Barchielli}}\ and\ \bibinfo {author} {\bibfnamefont {V~P}\ \bibnamefont
  {Belavkin}},\ }\bibfield  {title} {\enquote {\bibinfo {title} {Measurements
  continuous in time and a posteriori states in quantum mechanics},}\ }\href
  {\doibase 10.1088/0305-4470/24/7/022} {\bibfield  {journal} {\bibinfo
  {journal} {Journal of Physics A: Mathematical and General}\ }\textbf
  {\bibinfo {volume} {24}},\ \bibinfo {pages} {1495} (\bibinfo {year}
  {1991})}\BibitemShut {NoStop}%
\bibitem [{\citenamefont {Brun}(2002)}]{Brun2002}%
  \BibitemOpen
  \bibfield  {author} {\bibinfo {author} {\bibfnamefont {Todd~A.}\ \bibnamefont
  {Brun}},\ }\bibfield  {title} {\enquote {\bibinfo {title} {{A simple model of
  quantum trajectories}},}\ }\href {\doibase 10.1119/1.1475328} {\bibfield
  {journal} {\bibinfo  {journal} {American Journal of Physics}\ }\textbf
  {\bibinfo {volume} {70}},\ \bibinfo {pages} {719--737} (\bibinfo {year}
  {2002})}\BibitemShut {NoStop}%
\bibitem [{\citenamefont {Barchielli}(2006)}]{Bar06}%
  \BibitemOpen
  \bibfield  {author} {\bibinfo {author} {\bibfnamefont {Alberto}\ \bibnamefont
  {Barchielli}},\ }\enquote {\bibinfo {title} {Continual measurements in
  quantum mechanics and quantum stochastic calculus},}\ in\ \href {\doibase
  10.1007/3-540-33967-1_5} {\emph {\bibinfo {booktitle} {Open Quantum Systems
  III: Recent Developments}}},\ \bibinfo {editor} {edited by\ \bibinfo {editor}
  {\bibfnamefont {St{\'e}phane}\ \bibnamefont {Attal}}, \bibinfo {editor}
  {\bibfnamefont {Alain}\ \bibnamefont {Joye}}, \ and\ \bibinfo {editor}
  {\bibfnamefont {Claude-Alain}\ \bibnamefont {Pillet}}}\ (\bibinfo
  {publisher} {Springer Berlin Heidelberg},\ \bibinfo {address} {Berlin,
  Heidelberg},\ \bibinfo {year} {2006})\ pp.\ \bibinfo {pages}
  {207--292}\BibitemShut {NoStop}%
\bibitem [{\citenamefont {Belavkin}(1990)}]{Bel90}%
  \BibitemOpen
  \bibfield  {author} {\bibinfo {author} {\bibfnamefont {V.~P.}\ \bibnamefont
  {Belavkin}},\ }\bibfield  {title} {\enquote {\bibinfo {title} {A stochastic
  posterior schr{\"o}dinger equation for counting nondemolition measurement},}\
  }\href {\doibase 10.1007/BF00398273} {\bibfield  {journal} {\bibinfo
  {journal} {Letters in Mathematical Physics}\ }\textbf {\bibinfo {volume}
  {20}},\ \bibinfo {pages} {85--89} (\bibinfo {year} {1990})}\BibitemShut
  {NoStop}%
\bibitem [{\citenamefont {Belavkin}\ and\ \citenamefont
  {Staszewski}(1991)}]{BelStas91}%
  \BibitemOpen
  \bibfield  {author} {\bibinfo {author} {\bibfnamefont {V.P.}\ \bibnamefont
  {Belavkin}}\ and\ \bibinfo {author} {\bibfnamefont {P.}~\bibnamefont
  {Staszewski}},\ }\bibfield  {title} {\enquote {\bibinfo {title} {A continuous
  observation of photon emission},}\ }\href {\doibase
  https://doi.org/10.1016/0034-4877(91)90005-8} {\bibfield  {journal} {\bibinfo
   {journal} {Reports on Mathematical Physics}\ }\textbf {\bibinfo {volume}
  {29}},\ \bibinfo {pages} {213--225} (\bibinfo {year} {1991})}\BibitemShut
  {NoStop}%
\bibitem [{\citenamefont {Ghirardi}\ \emph {et~al.}(1990)\citenamefont
  {Ghirardi}, \citenamefont {Pearle},\ and\ \citenamefont {Rimini}}]{GPR90}%
  \BibitemOpen
  \bibfield  {author} {\bibinfo {author} {\bibfnamefont {Gian~Carlo}\
  \bibnamefont {Ghirardi}}, \bibinfo {author} {\bibfnamefont {Philip}\
  \bibnamefont {Pearle}}, \ and\ \bibinfo {author} {\bibfnamefont {Alberto}\
  \bibnamefont {Rimini}},\ }\bibfield  {title} {\enquote {\bibinfo {title}
  {Markov processes in hilbert space and continuous spontaneous localization of
  systems of identical particles},}\ }\href {\doibase 10.1103/PhysRevA.42.78}
  {\bibfield  {journal} {\bibinfo  {journal} {Phys. Rev. A}\ }\textbf {\bibinfo
  {volume} {42}},\ \bibinfo {pages} {78--89} (\bibinfo {year}
  {1990})}\BibitemShut {NoStop}%
\bibitem [{\citenamefont {Ghirardi}\ \emph {et~al.}(1986)\citenamefont
  {Ghirardi}, \citenamefont {Rimini},\ and\ \citenamefont {Weber}}]{GRW85}%
  \BibitemOpen
  \bibfield  {author} {\bibinfo {author} {\bibfnamefont {G.~C.}\ \bibnamefont
  {Ghirardi}}, \bibinfo {author} {\bibfnamefont {A.}~\bibnamefont {Rimini}}, \
  and\ \bibinfo {author} {\bibfnamefont {T.}~\bibnamefont {Weber}},\ }\bibfield
   {title} {\enquote {\bibinfo {title} {Unified dynamics for microscopic and
  macroscopic systems},}\ }\href {\doibase 10.1103/PhysRevD.34.470} {\bibfield
  {journal} {\bibinfo  {journal} {Phys. Rev. D}\ }\textbf {\bibinfo {volume}
  {34}},\ \bibinfo {pages} {470--491} (\bibinfo {year} {1986})}\BibitemShut
  {NoStop}%
\bibitem [{\citenamefont {Davies}(1976)}]{Davies76}%
  \BibitemOpen
  \bibfield  {author} {\bibinfo {author} {\bibfnamefont {E.B.}\ \bibnamefont
  {Davies}},\ }\href@noop {} {\emph {\bibinfo {title} {Quantum Theory of Open
  Systems}}}\ (\bibinfo  {publisher} {Academic Press},\ \bibinfo {year}
  {1976})\BibitemShut {NoStop}%
\bibitem [{\citenamefont {Srinivas}\ and\ \citenamefont
  {Davies}(1981)}]{SriDa}%
  \BibitemOpen
  \bibfield  {author} {\bibinfo {author} {\bibfnamefont {M.~D.}\ \bibnamefont
  {Srinivas}}\ and\ \bibinfo {author} {\bibfnamefont {E.~B.}\ \bibnamefont
  {Davies}},\ }\bibfield  {title} {\enquote {\bibinfo {title} {Photon counting
  probabilities in quantum optics},}\ }\href {\doibase 10.1080/713820643}
  {\bibfield  {journal} {\bibinfo  {journal} {Optica Acta: International
  Journal of Optics}\ }\textbf {\bibinfo {volume} {28}},\ \bibinfo {pages}
  {981--996} (\bibinfo {year} {1981})}\BibitemShut {NoStop}%
\bibitem [{\citenamefont {Zoller}\ \emph {et~al.}(1987)\citenamefont {Zoller},
  \citenamefont {Marte},\ and\ \citenamefont {Walls}}]{ZoWaMa}%
  \BibitemOpen
  \bibfield  {author} {\bibinfo {author} {\bibfnamefont {P.}~\bibnamefont
  {Zoller}}, \bibinfo {author} {\bibfnamefont {M.}~\bibnamefont {Marte}}, \
  and\ \bibinfo {author} {\bibfnamefont {D.~F.}\ \bibnamefont {Walls}},\
  }\bibfield  {title} {\enquote {\bibinfo {title} {Quantum jumps in atomic
  systems},}\ }\href {\doibase 10.1103/PhysRevA.35.198} {\bibfield  {journal}
  {\bibinfo  {journal} {Phys. Rev. A}\ }\textbf {\bibinfo {volume} {35}},\
  \bibinfo {pages} {198--207} (\bibinfo {year} {1987})}\BibitemShut {NoStop}%
\bibitem [{\citenamefont {Dum}\ \emph {et~al.}(1992)\citenamefont {Dum},
  \citenamefont {Zoller},\ and\ \citenamefont {Ritsch}}]{Dum92}%
  \BibitemOpen
  \bibfield  {author} {\bibinfo {author} {\bibfnamefont {R.}~\bibnamefont
  {Dum}}, \bibinfo {author} {\bibfnamefont {P.}~\bibnamefont {Zoller}}, \ and\
  \bibinfo {author} {\bibfnamefont {H.}~\bibnamefont {Ritsch}},\ }\bibfield
  {title} {\enquote {\bibinfo {title} {Monte carlo simulation of the atomic
  master equation for spontaneous emission},}\ }\href {\doibase
  10.1103/PhysRevA.45.4879} {\bibfield  {journal} {\bibinfo  {journal} {Phys.
  Rev. A}\ }\textbf {\bibinfo {volume} {45}},\ \bibinfo {pages} {4879--4887}
  (\bibinfo {year} {1992})}\BibitemShut {NoStop}%
\bibitem [{\citenamefont {Gisin}\ and\ \citenamefont {Percival}(1992)}]{GP92}%
  \BibitemOpen
  \bibfield  {author} {\bibinfo {author} {\bibfnamefont {N}~\bibnamefont
  {Gisin}}\ and\ \bibinfo {author} {\bibfnamefont {I~C}\ \bibnamefont
  {Percival}},\ }\bibfield  {title} {\enquote {\bibinfo {title} {The
  quantum-state diffusion model applied to open systems},}\ }\href {\doibase
  10.1088/0305-4470/25/21/023} {\bibfield  {journal} {\bibinfo  {journal}
  {Journal of Physics A: Mathematical and General}\ }\textbf {\bibinfo {volume}
  {25}},\ \bibinfo {pages} {5677} (\bibinfo {year} {1992})}\BibitemShut
  {NoStop}%
\bibitem [{\citenamefont {Carmichael}(1999)}]{Carmichael1999}%
  \BibitemOpen
  \bibfield  {author} {\bibinfo {author} {\bibfnamefont {H.~J.}\ \bibnamefont
  {Carmichael}},\ }\bibfield  {title} {\enquote {\bibinfo {title} {Quantum
  jumps revisited: An overview of quantum trajectory theory},}\ }in\ \href@noop
  {} {\emph {\bibinfo {booktitle} {Quantum Future From Volta and Como to the
  Present and Beyond}}},\ \bibinfo {editor} {edited by\ \bibinfo {editor}
  {\bibfnamefont {Philippe}\ \bibnamefont {Blanchard}}\ and\ \bibinfo {editor}
  {\bibfnamefont {Arkadiusz}\ \bibnamefont {Jadczyk}}}\ (\bibinfo  {publisher}
  {Springer Berlin Heidelberg},\ \bibinfo {address} {Berlin, Heidelberg},\
  \bibinfo {year} {1999})\ pp.\ \bibinfo {pages} {15--36}\BibitemShut {NoStop}%
\bibitem [{\citenamefont {Durrett}(2019)}]{durrett19}%
  \BibitemOpen
  \bibfield  {author} {\bibinfo {author} {\bibfnamefont {Rick}\ \bibnamefont
  {Durrett}},\ }\href {\doibase 10.1017/9781108591034} {\emph {\bibinfo {title}
  {Probability: Theory and Examples}}},\ \bibinfo {edition} {5th}\ ed.,\
  Cambridge Series in Statistical and Probabilistic Mathematics\ (\bibinfo
  {publisher} {Cambridge University Press},\ \bibinfo {year}
  {2019})\BibitemShut {NoStop}%
\bibitem [{\citenamefont {Wiseman}\ and\ \citenamefont
  {Milburn}(1993)}]{Wiseman93}%
  \BibitemOpen
  \bibfield  {author} {\bibinfo {author} {\bibfnamefont {H.~M.}\ \bibnamefont
  {Wiseman}}\ and\ \bibinfo {author} {\bibfnamefont {G.~J.}\ \bibnamefont
  {Milburn}},\ }\bibfield  {title} {\enquote {\bibinfo {title} {Quantum theory
  of field-quadrature measurements},}\ }\href {\doibase
  10.1103/PhysRevA.47.642} {\bibfield  {journal} {\bibinfo  {journal} {Phys.
  Rev. A}\ }\textbf {\bibinfo {volume} {47}},\ \bibinfo {pages} {642--662}
  (\bibinfo {year} {1993})}\BibitemShut {NoStop}%
\bibitem [{\citenamefont {Nha}\ and\ \citenamefont
  {Carmichael}(2004)}]{Nha2004}%
  \BibitemOpen
  \bibfield  {author} {\bibinfo {author} {\bibfnamefont {Hyunchul}\
  \bibnamefont {Nha}}\ and\ \bibinfo {author} {\bibfnamefont {H.~J.}\
  \bibnamefont {Carmichael}},\ }\bibfield  {title} {\enquote {\bibinfo {title}
  {Entanglement within the quantum trajectory description of open quantum
  systems},}\ }\href {\doibase 10.1103/PhysRevLett.93.120408} {\bibfield
  {journal} {\bibinfo  {journal} {Phys. Rev. Lett.}\ }\textbf {\bibinfo
  {volume} {93}},\ \bibinfo {pages} {120408} (\bibinfo {year}
  {2004})}\BibitemShut {NoStop}%
\bibitem [{\citenamefont {Bruno}\ \emph {et~al.}(2019)\citenamefont {Bruno},
  \citenamefont {Bianchet}, \citenamefont {Prakash}, \citenamefont {Li},
  \citenamefont {Alves},\ and\ \citenamefont {Mitchell}}]{Bruno2019}%
  \BibitemOpen
  \bibfield  {author} {\bibinfo {author} {\bibfnamefont {Natalia}\ \bibnamefont
  {Bruno}}, \bibinfo {author} {\bibfnamefont {Lorena~C.}\ \bibnamefont
  {Bianchet}}, \bibinfo {author} {\bibfnamefont {Vindhiya}\ \bibnamefont
  {Prakash}}, \bibinfo {author} {\bibfnamefont {Nan}\ \bibnamefont {Li}},
  \bibinfo {author} {\bibfnamefont {Nat\'{a}lia}\ \bibnamefont {Alves}}, \ and\
  \bibinfo {author} {\bibfnamefont {Morgan~W.}\ \bibnamefont {Mitchell}},\
  }\bibfield  {title} {\enquote {\bibinfo {title} {Maltese cross coupling to
  individual cold atoms in free space},}\ }\href {\doibase
  10.1364/OE.27.031042} {\bibfield  {journal} {\bibinfo  {journal} {Opt.
  Express}\ }\textbf {\bibinfo {volume} {27}},\ \bibinfo {pages} {31042--31052}
  (\bibinfo {year} {2019})}\BibitemShut {NoStop}%
\bibitem [{\citenamefont {Weber}\ \emph {et~al.}(2006)\citenamefont {Weber},
  \citenamefont {Volz}, \citenamefont {Saucke}, \citenamefont {Kurtsiefer},\
  and\ \citenamefont {Weinfurter}}]{Markus2006}%
  \BibitemOpen
  \bibfield  {author} {\bibinfo {author} {\bibfnamefont {Markus}\ \bibnamefont
  {Weber}}, \bibinfo {author} {\bibfnamefont {J\"urgen}\ \bibnamefont {Volz}},
  \bibinfo {author} {\bibfnamefont {Karen}\ \bibnamefont {Saucke}}, \bibinfo
  {author} {\bibfnamefont {Christian}\ \bibnamefont {Kurtsiefer}}, \ and\
  \bibinfo {author} {\bibfnamefont {Harald}\ \bibnamefont {Weinfurter}},\
  }\bibfield  {title} {\enquote {\bibinfo {title} {Analysis of a single-atom
  dipole trap},}\ }\href {\doibase 10.1103/PhysRevA.73.043406} {\bibfield
  {journal} {\bibinfo  {journal} {Phys. Rev. A}\ }\textbf {\bibinfo {volume}
  {73}},\ \bibinfo {pages} {043406} (\bibinfo {year} {2006})}\BibitemShut
  {NoStop}%
\bibitem [{SM()}]{SM}%
  \BibitemOpen
  \href@noop {} {}\bibinfo {note} {{s}ee Supplementary Material}\BibitemShut
  {NoStop}%
\bibitem [{\citenamefont {Mollow}(1975)}]{Mollow1975}%
  \BibitemOpen
  \bibfield  {author} {\bibinfo {author} {\bibfnamefont {B.~R.}\ \bibnamefont
  {Mollow}},\ }\bibfield  {title} {\enquote {\bibinfo {title} {Pure-state
  analysis of resonant light scattering: Radiative damping, saturation, and
  multiphoton effects},}\ }\href {\doibase 10.1103/PhysRevA.12.1919} {\bibfield
   {journal} {\bibinfo  {journal} {Phys. Rev. A}\ }\textbf {\bibinfo {volume}
  {12}},\ \bibinfo {pages} {1919--1943} (\bibinfo {year} {1975})}\BibitemShut
  {NoStop}%
\bibitem [{\citenamefont {Cook}(1980)}]{Cook1980}%
  \BibitemOpen
  \bibfield  {author} {\bibinfo {author} {\bibfnamefont {Richard~J.}\
  \bibnamefont {Cook}},\ }\bibfield  {title} {\enquote {\bibinfo {title}
  {Photon statistics in resonance fluorescence from laser deflection of an
  atomic beam},}\ }\href {\doibase
  https://doi.org/10.1016/0030-4018(80)90048-6} {\bibfield  {journal} {\bibinfo
   {journal} {Optics Communications}\ }\textbf {\bibinfo {volume} {35}},\
  \bibinfo {pages} {347--350} (\bibinfo {year} {1980})}\BibitemShut {NoStop}%
\bibitem [{\citenamefont {Cook}(1981)}]{Cook1981}%
  \BibitemOpen
  \bibfield  {author} {\bibinfo {author} {\bibfnamefont {Richard~J.}\
  \bibnamefont {Cook}},\ }\bibfield  {title} {\enquote {\bibinfo {title}
  {Photon number statistics in resonance fluorescence},}\ }\href {\doibase
  10.1103/PhysRevA.23.1243} {\bibfield  {journal} {\bibinfo  {journal} {Phys.
  Rev. A}\ }\textbf {\bibinfo {volume} {23}},\ \bibinfo {pages} {1243--1250}
  (\bibinfo {year} {1981})}\BibitemShut {NoStop}%
\bibitem [{\citenamefont {Carmichael}\ \emph {et~al.}(1989)\citenamefont
  {Carmichael}, \citenamefont {Singh}, \citenamefont {Vyas},\ and\
  \citenamefont {Rice}}]{Carmichael1989}%
  \BibitemOpen
  \bibfield  {author} {\bibinfo {author} {\bibfnamefont {H.~J.}\ \bibnamefont
  {Carmichael}}, \bibinfo {author} {\bibfnamefont {Surendra}\ \bibnamefont
  {Singh}}, \bibinfo {author} {\bibfnamefont {Reeta}\ \bibnamefont {Vyas}}, \
  and\ \bibinfo {author} {\bibfnamefont {P.~R.}\ \bibnamefont {Rice}},\
  }\bibfield  {title} {\enquote {\bibinfo {title} {Photoelectron waiting times
  and atomic state reduction in resonance fluorescence},}\ }\href {\doibase
  10.1103/PhysRevA.39.1200} {\bibfield  {journal} {\bibinfo  {journal} {Phys.
  Rev. A}\ }\textbf {\bibinfo {volume} {39}},\ \bibinfo {pages} {1200--1218}
  (\bibinfo {year} {1989})}\BibitemShut {NoStop}%
\bibitem [{\citenamefont {Walls}\ and\ \citenamefont
  {Zoller}(1981)}]{WallsZoller1981}%
  \BibitemOpen
  \bibfield  {author} {\bibinfo {author} {\bibfnamefont {D.~F.}\ \bibnamefont
  {Walls}}\ and\ \bibinfo {author} {\bibfnamefont {P.}~\bibnamefont {Zoller}},\
  }\bibfield  {title} {\enquote {\bibinfo {title} {Reduced quantum fluctuations
  in resonance fluorescence},}\ }\href {\doibase 10.1103/PhysRevLett.47.709}
  {\bibfield  {journal} {\bibinfo  {journal} {Phys. Rev. Lett.}\ }\textbf
  {\bibinfo {volume} {47}},\ \bibinfo {pages} {709--711} (\bibinfo {year}
  {1981})}\BibitemShut {NoStop}%
\bibitem [{\citenamefont {Mandel}(1982)}]{Mandel1982}%
  \BibitemOpen
  \bibfield  {author} {\bibinfo {author} {\bibfnamefont {L.}~\bibnamefont
  {Mandel}},\ }\bibfield  {title} {\enquote {\bibinfo {title} {Squeezed states
  and sub-poissonian photon statistics},}\ }\href {\doibase
  10.1103/PhysRevLett.49.136} {\bibfield  {journal} {\bibinfo  {journal} {Phys.
  Rev. Lett.}\ }\textbf {\bibinfo {volume} {49}},\ \bibinfo {pages} {136--138}
  (\bibinfo {year} {1982})}\BibitemShut {NoStop}%
\bibitem [{\citenamefont {Keys}(2022)}]{KeysPhD}%
  \BibitemOpen
  \bibfield  {author} {\bibinfo {author} {\bibfnamefont {Dustin~Michael}\
  \bibnamefont {Keys}},\ }\href {http://hdl.handle.net/10150/664319} {\enquote
  {\bibinfo {title} {A quantum stochastic approach to poisson master equation
  unravellings and ghirardi-rimini-weber theory},}\ } (\bibinfo {year}
  {2022})\BibitemShut {NoStop}%
\bibitem [{\citenamefont {Gisin}\ and\ \citenamefont
  {Percival}(1997)}]{gisin1997quantum}%
  \BibitemOpen
  \bibfield  {author} {\bibinfo {author} {\bibfnamefont {Nicolas}\ \bibnamefont
  {Gisin}}\ and\ \bibinfo {author} {\bibfnamefont {Ian~C}\ \bibnamefont
  {Percival}},\ }\bibfield  {title} {\enquote {\bibinfo {title} {Quantum state
  diffusion: from foundations to applications},}\ }\href {\doibase
  https://doi.org/10.48550/arXiv.quant-ph/9701024} {\  (\bibinfo {year}
  {1997}),\ https://doi.org/10.48550/arXiv.quant-ph/9701024}\BibitemShut
  {NoStop}%
\bibitem [{\citenamefont {Percival}(1998)}]{percival1998}%
  \BibitemOpen
  \bibfield  {author} {\bibinfo {author} {\bibfnamefont {Ian}\ \bibnamefont
  {Percival}},\ }\href@noop {} {\emph {\bibinfo {title} {Quantum State
  Diffusion}}}\ (\bibinfo  {publisher} {Cambridge University Press, Cambridge,
  UK},\ \bibinfo {year} {1998})\BibitemShut {NoStop}%
\bibitem [{\citenamefont {Natarajan}\ \emph {et~al.}(2012)\citenamefont
  {Natarajan}, \citenamefont {Tanner},\ and\ \citenamefont
  {Hadfield}}]{Natarajan_2012}%
  \BibitemOpen
  \bibfield  {author} {\bibinfo {author} {\bibfnamefont {Chandra~M}\
  \bibnamefont {Natarajan}}, \bibinfo {author} {\bibfnamefont {Michael~G}\
  \bibnamefont {Tanner}}, \ and\ \bibinfo {author} {\bibfnamefont {Robert~H}\
  \bibnamefont {Hadfield}},\ }\bibfield  {title} {\enquote {\bibinfo {title}
  {Superconducting nanowire single-photon detectors: physics and
  applications},}\ }\href {\doibase 10.1088/0953-2048/25/6/063001} {\bibfield
  {journal} {\bibinfo  {journal} {Superconductor Science and Technology}\
  }\textbf {\bibinfo {volume} {25}},\ \bibinfo {pages} {063001} (\bibinfo
  {year} {2012})}\BibitemShut {NoStop}%
\bibitem [{\citenamefont {Chou}\ \emph {et~al.}(2017)\citenamefont {Chou},
  \citenamefont {Auchter}, \citenamefont {Lilieholm}, \citenamefont {Smith},\
  and\ \citenamefont {Blinov}}]{Chou2017}%
  \BibitemOpen
  \bibfield  {author} {\bibinfo {author} {\bibfnamefont {Chen-Kuan}\
  \bibnamefont {Chou}}, \bibinfo {author} {\bibfnamefont {Carolyn}\
  \bibnamefont {Auchter}}, \bibinfo {author} {\bibfnamefont {Jennifer}\
  \bibnamefont {Lilieholm}}, \bibinfo {author} {\bibfnamefont {Kevin}\
  \bibnamefont {Smith}}, \ and\ \bibinfo {author} {\bibfnamefont {Boris}\
  \bibnamefont {Blinov}},\ }\bibfield  {title} {\enquote {\bibinfo {title}
  {{Note: Single ion imaging and fluorescence collection with a parabolic
  mirror trap}},}\ }\href {\doibase 10.1063/1.4996506} {\bibfield  {journal}
  {\bibinfo  {journal} {Review of Scientific Instruments}\ }\textbf {\bibinfo
  {volume} {88}},\ \bibinfo {pages} {086101} (\bibinfo {year}
  {2017})}\BibitemShut {NoStop}%
\bibitem [{\citenamefont {Jones}\ \emph {et~al.}(2007)\citenamefont {Jones},
  \citenamefont {Wiseman},\ and\ \citenamefont {Doherty}}]{Jones2007}%
  \BibitemOpen
  \bibfield  {author} {\bibinfo {author} {\bibfnamefont {S.~J.}\ \bibnamefont
  {Jones}}, \bibinfo {author} {\bibfnamefont {H.~M.}\ \bibnamefont {Wiseman}},
  \ and\ \bibinfo {author} {\bibfnamefont {A.~C.}\ \bibnamefont {Doherty}},\
  }\bibfield  {title} {\enquote {\bibinfo {title} {Entanglement,
  einstein-podolsky-rosen correlations, bell nonlocality, and steering},}\
  }\href {\doibase 10.1103/PhysRevA.76.052116} {\bibfield  {journal} {\bibinfo
  {journal} {Phys. Rev. A}\ }\textbf {\bibinfo {volume} {76}},\ \bibinfo
  {pages} {052116} (\bibinfo {year} {2007})}\BibitemShut {NoStop}%
\bibitem [{\citenamefont {Wiseman}\ and\ \citenamefont
  {Gambetta}(2012)}]{Wiseman2012}%
  \BibitemOpen
  \bibfield  {author} {\bibinfo {author} {\bibfnamefont {Howard~M.}\
  \bibnamefont {Wiseman}}\ and\ \bibinfo {author} {\bibfnamefont {Jay~M.}\
  \bibnamefont {Gambetta}},\ }\bibfield  {title} {\enquote {\bibinfo {title}
  {Are dynamical quantum jumps detector dependent?}}\ }\href {\doibase
  10.1103/PhysRevLett.108.220402} {\bibfield  {journal} {\bibinfo  {journal}
  {Phys. Rev. Lett.}\ }\textbf {\bibinfo {volume} {108}},\ \bibinfo {pages}
  {220402} (\bibinfo {year} {2012})}\BibitemShut {NoStop}%
\bibitem [{\citenamefont {Daryanoosh}\ and\ \citenamefont
  {Wiseman}(2014)}]{Daryanoosh_2014}%
  \BibitemOpen
  \bibfield  {author} {\bibinfo {author} {\bibfnamefont {Shakib}\ \bibnamefont
  {Daryanoosh}}\ and\ \bibinfo {author} {\bibfnamefont {Howard~M}\ \bibnamefont
  {Wiseman}},\ }\bibfield  {title} {\enquote {\bibinfo {title} {Quantum jumps
  are more quantum than quantum diffusion},}\ }\href {\doibase
  10.1088/1367-2630/16/6/063028} {\bibfield  {journal} {\bibinfo  {journal}
  {New Journal of Physics}\ }\textbf {\bibinfo {volume} {16}},\ \bibinfo
  {pages} {063028} (\bibinfo {year} {2014})}\BibitemShut {NoStop}%
\bibitem [{\citenamefont {Barchielli}\ and\ \citenamefont
  {Gregoratti}(2012)}]{BarchielliGregoratti2012}%
  \BibitemOpen
  \bibfield  {author} {\bibinfo {author} {\bibfnamefont {Alberto}\ \bibnamefont
  {Barchielli}}\ and\ \bibinfo {author} {\bibfnamefont {Matteo}\ \bibnamefont
  {Gregoratti}},\ }\bibfield  {title} {\enquote {\bibinfo {title} {Quantum
  measurements in continuous time, non-markovian evolutions and feedback},}\
  }\href {\doibase 10.1098/rsta.2011.0515} {\bibfield  {journal} {\bibinfo
  {journal} {Philosophical Transactions of the Royal Society A: Mathematical,
  Physical and Engineering Sciences}\ }\textbf {\bibinfo {volume} {370}},\
  \bibinfo {pages} {5364--5385} (\bibinfo {year} {2012})}\BibitemShut {NoStop}%
\bibitem [{\citenamefont {Arranz~Regidor}\ \emph {et~al.}(2021)\citenamefont
  {Arranz~Regidor}, \citenamefont {Crowder}, \citenamefont {Carmichael},\ and\
  \citenamefont {Hughes}}]{Arranz2021}%
  \BibitemOpen
  \bibfield  {author} {\bibinfo {author} {\bibfnamefont {Sofia}\ \bibnamefont
  {Arranz~Regidor}}, \bibinfo {author} {\bibfnamefont {Gavin}\ \bibnamefont
  {Crowder}}, \bibinfo {author} {\bibfnamefont {Howard}\ \bibnamefont
  {Carmichael}}, \ and\ \bibinfo {author} {\bibfnamefont {Stephen}\
  \bibnamefont {Hughes}},\ }\bibfield  {title} {\enquote {\bibinfo {title}
  {Modeling quantum light-matter interactions in waveguide qed with
  retardation, nonlinear interactions, and a time-delayed feedback: Matrix
  product states versus a space-discretized waveguide model},}\ }\href
  {\doibase 10.1103/PhysRevResearch.3.023030} {\bibfield  {journal} {\bibinfo
  {journal} {Phys. Rev. Res.}\ }\textbf {\bibinfo {volume} {3}},\ \bibinfo
  {pages} {023030} (\bibinfo {year} {2021})}\BibitemShut {NoStop}%
\bibitem [{\citenamefont {Preskill}(2018)}]{Preskill2018quantumcomputingin}%
  \BibitemOpen
  \bibfield  {author} {\bibinfo {author} {\bibfnamefont {John}\ \bibnamefont
  {Preskill}},\ }\bibfield  {title} {\enquote {\bibinfo {title} {Quantum
  {C}omputing in the {NISQ} era and beyond},}\ }\href {\doibase
  10.22331/q-2018-08-06-79} {\bibfield  {journal} {\bibinfo  {journal}
  {{Quantum}}\ }\textbf {\bibinfo {volume} {2}},\ \bibinfo {pages} {79}
  (\bibinfo {year} {2018})}\BibitemShut {NoStop}%
\bibitem [{\citenamefont {Altman}\ \emph {et~al.}(2021)\citenamefont {Altman},
  \citenamefont {Brown}, \citenamefont {Carleo}, \citenamefont {Carr},
  \citenamefont {Demler}, \citenamefont {Chin}, \citenamefont {DeMarco},
  \citenamefont {Economou}, \citenamefont {Eriksson}, \citenamefont {Fu},
  \citenamefont {Greiner}, \citenamefont {Hazzard}, \citenamefont {Hulet},
  \citenamefont {Koll\'ar}, \citenamefont {Lev}, \citenamefont {Lukin},
  \citenamefont {Ma}, \citenamefont {Mi}, \citenamefont {Misra}, \citenamefont
  {Monroe}, \citenamefont {Murch}, \citenamefont {Nazario}, \citenamefont {Ni},
  \citenamefont {Potter}, \citenamefont {Roushan}, \citenamefont {Saffman},
  \citenamefont {Schleier-Smith}, \citenamefont {Siddiqi}, \citenamefont
  {Simmonds}, \citenamefont {Singh}, \citenamefont {Spielman}, \citenamefont
  {Temme}, \citenamefont {Weiss}, \citenamefont {Vu\ifmmode \check{c}\else
  \v{c}\fi{}kovi\ifmmode~\acute{c}\else \'{c}\fi{}}, \citenamefont
  {Vuleti\ifmmode~\acute{c}\else \'{c}\fi{}}, \citenamefont {Ye},\ and\
  \citenamefont {Zwierlein}}]{Altman21}%
  \BibitemOpen
  \bibfield  {author} {\bibinfo {author} {\bibfnamefont {Ehud}\ \bibnamefont
  {Altman}}, \bibinfo {author} {\bibfnamefont {Kenneth~R.}\ \bibnamefont
  {Brown}}, \bibinfo {author} {\bibfnamefont {Giuseppe}\ \bibnamefont
  {Carleo}}, \bibinfo {author} {\bibfnamefont {Lincoln~D.}\ \bibnamefont
  {Carr}}, \bibinfo {author} {\bibfnamefont {Eugene}\ \bibnamefont {Demler}},
  \bibinfo {author} {\bibfnamefont {Cheng}\ \bibnamefont {Chin}}, \bibinfo
  {author} {\bibfnamefont {Brian}\ \bibnamefont {DeMarco}}, \bibinfo {author}
  {\bibfnamefont {Sophia~E.}\ \bibnamefont {Economou}}, \bibinfo {author}
  {\bibfnamefont {Mark~A.}\ \bibnamefont {Eriksson}}, \bibinfo {author}
  {\bibfnamefont {Kai-Mei~C.}\ \bibnamefont {Fu}}, \bibinfo {author}
  {\bibfnamefont {Markus}\ \bibnamefont {Greiner}}, \bibinfo {author}
  {\bibfnamefont {Kaden~R.A.}\ \bibnamefont {Hazzard}}, \bibinfo {author}
  {\bibfnamefont {Randall~G.}\ \bibnamefont {Hulet}}, \bibinfo {author}
  {\bibfnamefont {Alicia~J.}\ \bibnamefont {Koll\'ar}}, \bibinfo {author}
  {\bibfnamefont {Benjamin~L.}\ \bibnamefont {Lev}}, \bibinfo {author}
  {\bibfnamefont {Mikhail~D.}\ \bibnamefont {Lukin}}, \bibinfo {author}
  {\bibfnamefont {Ruichao}\ \bibnamefont {Ma}}, \bibinfo {author}
  {\bibfnamefont {Xiao}\ \bibnamefont {Mi}}, \bibinfo {author} {\bibfnamefont
  {Shashank}\ \bibnamefont {Misra}}, \bibinfo {author} {\bibfnamefont
  {Christopher}\ \bibnamefont {Monroe}}, \bibinfo {author} {\bibfnamefont
  {Kater}\ \bibnamefont {Murch}}, \bibinfo {author} {\bibfnamefont {Zaira}\
  \bibnamefont {Nazario}}, \bibinfo {author} {\bibfnamefont {Kang-Kuen}\
  \bibnamefont {Ni}}, \bibinfo {author} {\bibfnamefont {Andrew~C.}\
  \bibnamefont {Potter}}, \bibinfo {author} {\bibfnamefont {Pedram}\
  \bibnamefont {Roushan}}, \bibinfo {author} {\bibfnamefont {Mark}\
  \bibnamefont {Saffman}}, \bibinfo {author} {\bibfnamefont {Monika}\
  \bibnamefont {Schleier-Smith}}, \bibinfo {author} {\bibfnamefont {Irfan}\
  \bibnamefont {Siddiqi}}, \bibinfo {author} {\bibfnamefont {Raymond}\
  \bibnamefont {Simmonds}}, \bibinfo {author} {\bibfnamefont {Meenakshi}\
  \bibnamefont {Singh}}, \bibinfo {author} {\bibfnamefont {I.B.}\ \bibnamefont
  {Spielman}}, \bibinfo {author} {\bibfnamefont {Kristan}\ \bibnamefont
  {Temme}}, \bibinfo {author} {\bibfnamefont {David~S.}\ \bibnamefont {Weiss}},
  \bibinfo {author} {\bibfnamefont {Jelena}\ \bibnamefont {Vu\ifmmode
  \check{c}\else \v{c}\fi{}kovi\ifmmode~\acute{c}\else \'{c}\fi{}}}, \bibinfo
  {author} {\bibfnamefont {Vladan}\ \bibnamefont {Vuleti\ifmmode~\acute{c}\else
  \'{c}\fi{}}}, \bibinfo {author} {\bibfnamefont {Jun}\ \bibnamefont {Ye}}, \
  and\ \bibinfo {author} {\bibfnamefont {Martin}\ \bibnamefont {Zwierlein}},\
  }\bibfield  {title} {\enquote {\bibinfo {title} {Quantum simulators:
  Architectures and opportunities},}\ }\href {\doibase
  10.1103/PRXQuantum.2.017003} {\bibfield  {journal} {\bibinfo  {journal} {PRX
  Quantum}\ }\textbf {\bibinfo {volume} {2}},\ \bibinfo {pages} {017003}
  (\bibinfo {year} {2021})}\BibitemShut {NoStop}%
\bibitem [{\citenamefont {Fraxanet}\ \emph {et~al.}(2023)\citenamefont
  {Fraxanet}, \citenamefont {Salamon},\ and\ \citenamefont
  {Lewenstein}}]{fraxanet2022coming}%
  \BibitemOpen
  \bibfield  {author} {\bibinfo {author} {\bibfnamefont {Joana}\ \bibnamefont
  {Fraxanet}}, \bibinfo {author} {\bibfnamefont {Tymoteusz}\ \bibnamefont
  {Salamon}}, \ and\ \bibinfo {author} {\bibfnamefont {Maciej}\ \bibnamefont
  {Lewenstein}},\ }\enquote {\bibinfo {title} {The coming decades of quantum
  simulation},}\ in\ \href {\doibase 10.1007/978-3-031-32469-7_4} {\emph
  {\bibinfo {booktitle} {Sketches of Physics: The Celebration Collection}}},\
  \bibinfo {editor} {edited by\ \bibinfo {editor} {\bibfnamefont {Roberta}\
  \bibnamefont {Citro}}, \bibinfo {editor} {\bibfnamefont {Maciej}\
  \bibnamefont {Lewenstein}}, \bibinfo {editor} {\bibfnamefont {Angel}\
  \bibnamefont {Rubio}}, \bibinfo {editor} {\bibfnamefont {Wolfgang~P.}\
  \bibnamefont {Schleich}}, \bibinfo {editor} {\bibfnamefont {James~D.}\
  \bibnamefont {Wells}}, \ and\ \bibinfo {editor} {\bibfnamefont {Gary~P.}\
  \bibnamefont {Zank}}}\ (\bibinfo  {publisher} {Springer International
  Publishing},\ \bibinfo {year} {2023})\ pp.\ \bibinfo {pages}
  {85--125}\BibitemShut {NoStop}%
\bibitem [{\citenamefont {Skinner}\ \emph {et~al.}(2019)\citenamefont
  {Skinner}, \citenamefont {Ruhman},\ and\ \citenamefont
  {Nahum}}]{skinner2019measurementinducedphase}%
  \BibitemOpen
  \bibfield  {author} {\bibinfo {author} {\bibfnamefont {Brian}\ \bibnamefont
  {Skinner}}, \bibinfo {author} {\bibfnamefont {Jonathan}\ \bibnamefont
  {Ruhman}}, \ and\ \bibinfo {author} {\bibfnamefont {Adam}\ \bibnamefont
  {Nahum}},\ }\bibfield  {title} {\enquote {\bibinfo {title}
  {Measurement-induced phase transitions in the dynamics of entanglement},}\
  }\href {\doibase 10.1103/PhysRevX.9.031009} {\bibfield  {journal} {\bibinfo
  {journal} {Phys. Rev. X}\ }\textbf {\bibinfo {volume} {9}},\ \bibinfo {pages}
  {031009} (\bibinfo {year} {2019})}\BibitemShut {NoStop}%
\bibitem [{\citenamefont {Li}\ \emph {et~al.}(2018)\citenamefont {Li},
  \citenamefont {Chen},\ and\ \citenamefont
  {Fisher}}]{li2018quantumzenoeffect}%
  \BibitemOpen
  \bibfield  {author} {\bibinfo {author} {\bibfnamefont {Yaodong}\ \bibnamefont
  {Li}}, \bibinfo {author} {\bibfnamefont {Xiao}\ \bibnamefont {Chen}}, \ and\
  \bibinfo {author} {\bibfnamefont {Matthew P.~A.}\ \bibnamefont {Fisher}},\
  }\bibfield  {title} {\enquote {\bibinfo {title} {Quantum zeno effect and the
  many-body entanglement transition},}\ }\href {\doibase
  10.1103/PhysRevB.98.205136} {\bibfield  {journal} {\bibinfo  {journal} {Phys.
  Rev. B}\ }\textbf {\bibinfo {volume} {98}},\ \bibinfo {pages} {205136}
  (\bibinfo {year} {2018})}\BibitemShut {NoStop}%
\bibitem [{\citenamefont {Li}\ \emph {et~al.}(2019)\citenamefont {Li},
  \citenamefont {Chen},\ and\ \citenamefont
  {Fisher}}]{li2019measurementdrivenentanglement}%
  \BibitemOpen
  \bibfield  {author} {\bibinfo {author} {\bibfnamefont {Yaodong}\ \bibnamefont
  {Li}}, \bibinfo {author} {\bibfnamefont {Xiao}\ \bibnamefont {Chen}}, \ and\
  \bibinfo {author} {\bibfnamefont {Matthew P.~A.}\ \bibnamefont {Fisher}},\
  }\bibfield  {title} {\enquote {\bibinfo {title} {Measurement-driven
  entanglement transition in hybrid quantum circuits},}\ }\href {\doibase
  10.1103/PhysRevB.100.134306} {\bibfield  {journal} {\bibinfo  {journal}
  {Phys. Rev. B}\ }\textbf {\bibinfo {volume} {100}},\ \bibinfo {pages}
  {134306} (\bibinfo {year} {2019})}\BibitemShut {NoStop}%
\bibitem [{\citenamefont {Chan}\ \emph {et~al.}(2019)\citenamefont {Chan},
  \citenamefont {Nandkishore}, \citenamefont {Pretko},\ and\ \citenamefont
  {Smith}}]{chan2019unitaryprojective}%
  \BibitemOpen
  \bibfield  {author} {\bibinfo {author} {\bibfnamefont {Amos}\ \bibnamefont
  {Chan}}, \bibinfo {author} {\bibfnamefont {Rahul~M.}\ \bibnamefont
  {Nandkishore}}, \bibinfo {author} {\bibfnamefont {Michael}\ \bibnamefont
  {Pretko}}, \ and\ \bibinfo {author} {\bibfnamefont {Graeme}\ \bibnamefont
  {Smith}},\ }\bibfield  {title} {\enquote {\bibinfo {title}
  {Unitary-projective entanglement dynamics},}\ }\href {\doibase
  10.1103/PhysRevB.99.224307} {\bibfield  {journal} {\bibinfo  {journal} {Phys.
  Rev. B}\ }\textbf {\bibinfo {volume} {99}},\ \bibinfo {pages} {224307}
  (\bibinfo {year} {2019})}\BibitemShut {NoStop}%
\bibitem [{\citenamefont {Sierant}\ and\ \citenamefont
  {Turkeshi}(2022)}]{sierant2022universalbehaviorbeyond}%
  \BibitemOpen
  \bibfield  {author} {\bibinfo {author} {\bibfnamefont {Piotr}\ \bibnamefont
  {Sierant}}\ and\ \bibinfo {author} {\bibfnamefont {Xhek}\ \bibnamefont
  {Turkeshi}},\ }\bibfield  {title} {\enquote {\bibinfo {title} {Universal
  behavior beyond multifractality of wave functions at measurement-induced
  phase transitions},}\ }\href {\doibase 10.1103/PhysRevLett.128.130605}
  {\bibfield  {journal} {\bibinfo  {journal} {Phys. Rev. Lett.}\ }\textbf
  {\bibinfo {volume} {128}},\ \bibinfo {pages} {130605} (\bibinfo {year}
  {2022})}\BibitemShut {NoStop}%
\bibitem [{\citenamefont {Turkeshi}\ \emph {et~al.}(2020)\citenamefont
  {Turkeshi}, \citenamefont {Fazio},\ and\ \citenamefont
  {Dalmonte}}]{Turkeshi20_2d}%
  \BibitemOpen
  \bibfield  {author} {\bibinfo {author} {\bibfnamefont {Xhek}\ \bibnamefont
  {Turkeshi}}, \bibinfo {author} {\bibfnamefont {Rosario}\ \bibnamefont
  {Fazio}}, \ and\ \bibinfo {author} {\bibfnamefont {Marcello}\ \bibnamefont
  {Dalmonte}},\ }\bibfield  {title} {\enquote {\bibinfo {title}
  {Measurement-induced criticality in $(2+1)$-dimensional hybrid quantum
  circuits},}\ }\href {\doibase 10.1103/PhysRevB.102.014315} {\bibfield
  {journal} {\bibinfo  {journal} {Phys. Rev. B}\ }\textbf {\bibinfo {volume}
  {102}},\ \bibinfo {pages} {014315} (\bibinfo {year} {2020})}\BibitemShut
  {NoStop}%
\bibitem [{\citenamefont {Lunt}\ \emph {et~al.}(2021)\citenamefont {Lunt},
  \citenamefont {Szyniszewski},\ and\ \citenamefont
  {Pal}}]{lunt2021measurementinducedcriticality}%
  \BibitemOpen
  \bibfield  {author} {\bibinfo {author} {\bibfnamefont {Oliver}\ \bibnamefont
  {Lunt}}, \bibinfo {author} {\bibfnamefont {Marcin}\ \bibnamefont
  {Szyniszewski}}, \ and\ \bibinfo {author} {\bibfnamefont {Arijeet}\
  \bibnamefont {Pal}},\ }\bibfield  {title} {\enquote {\bibinfo {title}
  {Measurement-induced criticality and entanglement clusters: A study of
  one-dimensional and two-dimensional clifford circuits},}\ }\href {\doibase
  10.1103/PhysRevB.104.155111} {\bibfield  {journal} {\bibinfo  {journal}
  {Phys. Rev. B}\ }\textbf {\bibinfo {volume} {104}},\ \bibinfo {pages}
  {155111} (\bibinfo {year} {2021})}\BibitemShut {NoStop}%
\bibitem [{\citenamefont {Sierant}\ \emph {et~al.}(2022)\citenamefont
  {Sierant}, \citenamefont {Schir\`o}, \citenamefont {Lewenstein},\ and\
  \citenamefont {Turkeshi}}]{Sierant22_2d}%
  \BibitemOpen
  \bibfield  {author} {\bibinfo {author} {\bibfnamefont {Piotr}\ \bibnamefont
  {Sierant}}, \bibinfo {author} {\bibfnamefont {Marco}\ \bibnamefont
  {Schir\`o}}, \bibinfo {author} {\bibfnamefont {Maciej}\ \bibnamefont
  {Lewenstein}}, \ and\ \bibinfo {author} {\bibfnamefont {Xhek}\ \bibnamefont
  {Turkeshi}},\ }\bibfield  {title} {\enquote {\bibinfo {title}
  {Measurement-induced phase transitions in $(d+1)$-dimensional stabilizer
  circuits},}\ }\href {\doibase 10.1103/PhysRevB.106.214316} {\bibfield
  {journal} {\bibinfo  {journal} {Phys. Rev. B}\ }\textbf {\bibinfo {volume}
  {106}},\ \bibinfo {pages} {214316} (\bibinfo {year} {2022})}\BibitemShut
  {NoStop}%
\bibitem [{\citenamefont {Cao}\ \emph {et~al.}(2019)\citenamefont {Cao},
  \citenamefont {Tilloy},\ and\ \citenamefont {Luca}}]{cao2019entanglementina}%
  \BibitemOpen
  \bibfield  {author} {\bibinfo {author} {\bibfnamefont {Xiangyu}\ \bibnamefont
  {Cao}}, \bibinfo {author} {\bibfnamefont {Antoine}\ \bibnamefont {Tilloy}}, \
  and\ \bibinfo {author} {\bibfnamefont {Andrea~De}\ \bibnamefont {Luca}},\
  }\bibfield  {title} {\enquote {\bibinfo {title} {{Entanglement in a fermion
  chain under continuous monitoring}},}\ }\href {\doibase
  10.21468/SciPostPhys.7.2.024} {\bibfield  {journal} {\bibinfo  {journal}
  {SciPost Phys.}\ }\textbf {\bibinfo {volume} {7}},\ \bibinfo {pages} {024}
  (\bibinfo {year} {2019})}\BibitemShut {NoStop}%
\bibitem [{\citenamefont {Piccitto}\ \emph {et~al.}(2022)\citenamefont
  {Piccitto}, \citenamefont {Russomanno},\ and\ \citenamefont
  {Rossini}}]{Piccitto2022}%
  \BibitemOpen
  \bibfield  {author} {\bibinfo {author} {\bibfnamefont {Giulia}\ \bibnamefont
  {Piccitto}}, \bibinfo {author} {\bibfnamefont {Angelo}\ \bibnamefont
  {Russomanno}}, \ and\ \bibinfo {author} {\bibfnamefont {Davide}\ \bibnamefont
  {Rossini}},\ }\bibfield  {title} {\enquote {\bibinfo {title} {Entanglement
  transitions in the quantum ising chain: A comparison between different
  unravelings of the same lindbladian},}\ }\href {\doibase
  10.1103/PhysRevB.105.064305} {\bibfield  {journal} {\bibinfo  {journal}
  {Phys. Rev. B}\ }\textbf {\bibinfo {volume} {105}},\ \bibinfo {pages}
  {064305} (\bibinfo {year} {2022})}\BibitemShut {NoStop}%
\bibitem [{\citenamefont {Kolodrubetz}(2023)}]{Kolodrubetz2021}%
  \BibitemOpen
  \bibfield  {author} {\bibinfo {author} {\bibfnamefont {Michael}\ \bibnamefont
  {Kolodrubetz}},\ }\bibfield  {title} {\enquote {\bibinfo {title} {Optimality
  of lindblad unfolding in measurement phase transitions},}\ }\href {\doibase
  10.1103/PhysRevB.107.L140301} {\bibfield  {journal} {\bibinfo  {journal}
  {Phys. Rev. B}\ }\textbf {\bibinfo {volume} {107}},\ \bibinfo {pages}
  {L140301} (\bibinfo {year} {2023})}\BibitemShut {NoStop}%
\bibitem [{\citenamefont {Turkeshi}\ \emph {et~al.}(2021)\citenamefont
  {Turkeshi}, \citenamefont {Biella}, \citenamefont {Fazio}, \citenamefont
  {Dalmonte},\ and\ \citenamefont {Schir\'o}}]{Turkeshi21clicks}%
  \BibitemOpen
  \bibfield  {author} {\bibinfo {author} {\bibfnamefont {Xhek}\ \bibnamefont
  {Turkeshi}}, \bibinfo {author} {\bibfnamefont {Alberto}\ \bibnamefont
  {Biella}}, \bibinfo {author} {\bibfnamefont {Rosario}\ \bibnamefont {Fazio}},
  \bibinfo {author} {\bibfnamefont {Marcello}\ \bibnamefont {Dalmonte}}, \ and\
  \bibinfo {author} {\bibfnamefont {Marco}\ \bibnamefont {Schir\'o}},\
  }\bibfield  {title} {\enquote {\bibinfo {title} {Measurement-induced
  entanglement transitions in the quantum ising chain: From infinite to zero
  clicks},}\ }\href {\doibase 10.1103/PhysRevB.103.224210} {\bibfield
  {journal} {\bibinfo  {journal} {Phys. Rev. B}\ }\textbf {\bibinfo {volume}
  {103}},\ \bibinfo {pages} {224210} (\bibinfo {year} {2021})}\BibitemShut
  {NoStop}%
\bibitem [{\citenamefont {Sierant}\ and\ \citenamefont
  {Turkeshi}(2023{\natexlab{a}})}]{Sierant23control}%
  \BibitemOpen
  \bibfield  {author} {\bibinfo {author} {\bibfnamefont {Piotr}\ \bibnamefont
  {Sierant}}\ and\ \bibinfo {author} {\bibfnamefont {Xhek}\ \bibnamefont
  {Turkeshi}},\ }\bibfield  {title} {\enquote {\bibinfo {title} {Controlling
  entanglement at absorbing state phase transitions in random circuits},}\
  }\href {\doibase 10.1103/PhysRevLett.130.120402} {\bibfield  {journal}
  {\bibinfo  {journal} {Phys. Rev. Lett.}\ }\textbf {\bibinfo {volume} {130}},\
  \bibinfo {pages} {120402} (\bibinfo {year} {2023}{\natexlab{a}})}\BibitemShut
  {NoStop}%
\bibitem [{\citenamefont {Buchhold}\ \emph {et~al.}()\citenamefont {Buchhold},
  \citenamefont {M\"uller},\ and\ \citenamefont
  {Diehl}}]{buchhold2022revealingmeasurementinduced}%
  \BibitemOpen
  \bibfield  {author} {\bibinfo {author} {\bibfnamefont {M.}~\bibnamefont
  {Buchhold}}, \bibinfo {author} {\bibfnamefont {T.}~\bibnamefont {M\"uller}},
  \ and\ \bibinfo {author} {\bibfnamefont {S.}~\bibnamefont {Diehl}},\
  }\href@noop {} {}\Eprint {http://arxiv.org/abs/2208.10506} {arXiv:2208.10506}
  \BibitemShut {NoStop}%
\bibitem [{\citenamefont {Iadecola}\ \emph {et~al.}(2023)\citenamefont
  {Iadecola}, \citenamefont {Ganeshan}, \citenamefont {Pixley},\ and\
  \citenamefont {Wilson}}]{iadecola2022dynamicalentanglementtransition}%
  \BibitemOpen
  \bibfield  {author} {\bibinfo {author} {\bibfnamefont {Thomas}\ \bibnamefont
  {Iadecola}}, \bibinfo {author} {\bibfnamefont {Sriram}\ \bibnamefont
  {Ganeshan}}, \bibinfo {author} {\bibfnamefont {J.~H.}\ \bibnamefont
  {Pixley}}, \ and\ \bibinfo {author} {\bibfnamefont {Justin~H.}\ \bibnamefont
  {Wilson}},\ }\bibfield  {title} {\enquote {\bibinfo {title} {Measurement and
  feedback driven entanglement transition in the probabilistic control of
  chaos},}\ }\href {\doibase 10.1103/PhysRevLett.131.060403} {\bibfield
  {journal} {\bibinfo  {journal} {Phys. Rev. Lett.}\ }\textbf {\bibinfo
  {volume} {131}},\ \bibinfo {pages} {060403} (\bibinfo {year}
  {2023})}\BibitemShut {NoStop}%
\bibitem [{\citenamefont {Piroli}\ \emph {et~al.}(2023)\citenamefont {Piroli},
  \citenamefont {Li}, \citenamefont {Vasseur},\ and\ \citenamefont
  {Nahum}}]{Piroli23triviality}%
  \BibitemOpen
  \bibfield  {author} {\bibinfo {author} {\bibfnamefont {Lorenzo}\ \bibnamefont
  {Piroli}}, \bibinfo {author} {\bibfnamefont {Yaodong}\ \bibnamefont {Li}},
  \bibinfo {author} {\bibfnamefont {Romain}\ \bibnamefont {Vasseur}}, \ and\
  \bibinfo {author} {\bibfnamefont {Adam}\ \bibnamefont {Nahum}},\ }\bibfield
  {title} {\enquote {\bibinfo {title} {Triviality of quantum trajectories close
  to a directed percolation transition},}\ }\href {\doibase
  10.1103/PhysRevB.107.224303} {\bibfield  {journal} {\bibinfo  {journal}
  {Phys. Rev. B}\ }\textbf {\bibinfo {volume} {107}},\ \bibinfo {pages}
  {224303} (\bibinfo {year} {2023})}\BibitemShut {NoStop}%
\bibitem [{\citenamefont {O'Dea}\ \emph {et~al.}(2024)\citenamefont {O'Dea},
  \citenamefont {Morningstar}, \citenamefont {Gopalakrishnan},\ and\
  \citenamefont {Khemani}}]{odea2022entanglementandabsorbing}%
  \BibitemOpen
  \bibfield  {author} {\bibinfo {author} {\bibfnamefont {Nicholas}\
  \bibnamefont {O'Dea}}, \bibinfo {author} {\bibfnamefont {Alan}\ \bibnamefont
  {Morningstar}}, \bibinfo {author} {\bibfnamefont {Sarang}\ \bibnamefont
  {Gopalakrishnan}}, \ and\ \bibinfo {author} {\bibfnamefont {Vedika}\
  \bibnamefont {Khemani}},\ }\bibfield  {title} {\enquote {\bibinfo {title}
  {Entanglement and absorbing-state transitions in interactive quantum
  dynamics},}\ }\href {\doibase 10.1103/PhysRevB.109.L020304} {\bibfield
  {journal} {\bibinfo  {journal} {Phys. Rev. B}\ }\textbf {\bibinfo {volume}
  {109}},\ \bibinfo {pages} {L020304} (\bibinfo {year} {2024})}\BibitemShut
  {NoStop}%
\bibitem [{\citenamefont {Ravindranath}\ \emph
  {et~al.}(2023{\natexlab{a}})\citenamefont {Ravindranath}, \citenamefont
  {Han}, \citenamefont {Yang},\ and\ \citenamefont
  {Chen}}]{ravindranath2022entanglementsteeringin}%
  \BibitemOpen
  \bibfield  {author} {\bibinfo {author} {\bibfnamefont {Vikram}\ \bibnamefont
  {Ravindranath}}, \bibinfo {author} {\bibfnamefont {Yiqiu}\ \bibnamefont
  {Han}}, \bibinfo {author} {\bibfnamefont {Zhi-Cheng}\ \bibnamefont {Yang}}, \
  and\ \bibinfo {author} {\bibfnamefont {Xiao}\ \bibnamefont {Chen}},\
  }\bibfield  {title} {\enquote {\bibinfo {title} {Entanglement steering in
  adaptive circuits with feedback},}\ }\href {\doibase
  10.1103/PhysRevB.108.L041103} {\bibfield  {journal} {\bibinfo  {journal}
  {Phys. Rev. B}\ }\textbf {\bibinfo {volume} {108}},\ \bibinfo {pages}
  {L041103} (\bibinfo {year} {2023}{\natexlab{a}})}\BibitemShut {NoStop}%
\bibitem [{\citenamefont {Ravindranath}\ \emph
  {et~al.}(2023{\natexlab{b}})\citenamefont {Ravindranath}, \citenamefont
  {Yang},\ and\ \citenamefont {Chen}}]{ravindranath2023free}%
  \BibitemOpen
  \bibfield  {author} {\bibinfo {author} {\bibfnamefont {Vikram}\ \bibnamefont
  {Ravindranath}}, \bibinfo {author} {\bibfnamefont {Zhi-Cheng}\ \bibnamefont
  {Yang}}, \ and\ \bibinfo {author} {\bibfnamefont {Xiao}\ \bibnamefont
  {Chen}},\ }\href@noop {} {\enquote {\bibinfo {title} {Free fermions under
  adaptive quantum dynamics},}\ } (\bibinfo {year} {2023}{\natexlab{b}}),\
  \Eprint {http://arxiv.org/abs/2306.16595} {arXiv:2306.16595 [quant-ph]}
  \BibitemShut {NoStop}%
\bibitem [{\citenamefont {Sierant}\ and\ \citenamefont
  {Turkeshi}(2023{\natexlab{b}})}]{sierant2023entanglement}%
  \BibitemOpen
  \bibfield  {author} {\bibinfo {author} {\bibfnamefont {P.}~\bibnamefont
  {Sierant}}\ and\ \bibinfo {author} {\bibfnamefont {X.}~\bibnamefont
  {Turkeshi}},\ }\bibfield  {title} {\enquote {\bibinfo {title} {Entanglement
  and absorbing state transitions in (d+1)-dimensional stabilizer circuits},}\
  }\href {\doibase 10.12693/APhysPolA.144.474} {\bibfield  {journal} {\bibinfo
  {journal} {Acta Physica Polonica A ISSN 1898-794X}\ }\textbf {\bibinfo
  {volume} {144}},\ \bibinfo {pages} {474} (\bibinfo {year}
  {2023}{\natexlab{b}})}\BibitemShut {NoStop}%
\end{thebibliography}%
\end{document}